\documentclass[10pt,journal,compsoc]{IEEEtran}
\usepackage{amsmath}
\ifCLASSOPTIONcompsoc
  \usepackage[caption=false,font=footnotesize,labelfont=sf,textfont=sf]{subfig}
\else
  \usepackage[caption=false,font=footnotesize]{subfig}
\fi
\usepackage{graphicx}
\usepackage{amssymb}
\usepackage{amsthm}
\usepackage{algorithm}
\usepackage[noend]{algorithmic}
\usepackage{xcolor}
\usepackage{url}
\usepackage{booktabs}
\usepackage{array}
\usepackage{tabularx}
\usepackage{placeins}
\ifCLASSOPTIONcompsoc
  \usepackage[nocompress]{cite}
\else
  \usepackage{cite}
\fi

\newcolumntype{Y}{>{\centering\arraybackslash}X}
\newcolumntype{L}{>{\raggedright\arraybackslash}X}

\newtheorem{definition}{Definition}
\newtheorem{theorem}{Theorem}
\newtheorem{lemma}{Lemma}
\newtheorem{proposition}{Proposition}
\newtheorem{corollary}{Corollary}
\newtheorem{remark}{Remark}
\newtheorem{assumption}{Assumption}
\newcommand{\ActSet}{\mathcal{A}}
\newcommand{\FrameLen}{F}
\newcommand{\policyvk}{\pi^{v,\FrameLen}}
\newcommand{\Qhat}{\widehat{Q}}
\newcommand{\framehead}{\tau_n}
\newcommand{\tabnote}[1]{\vspace{2pt}\par\raggedright\scriptsize #1\par}

\makeatletter
\@namedef{r@app:capacity}{{A}{0}}
\@namedef{r@app:virtual}{{B}{0}}
\@namedef{r@app:physical}{{C}{0}}
\@namedef{r@app:virtual_interval}{{C.1}{0}}
\@namedef{r@app:physical_arrival}{{C.2}{0}}
\@namedef{r@app:waiting_bound}{{C.3}{0}}
\@namedef{r@app:nto_induction}{{C.4}{0}}
\@namedef{r@app:throughput_proof}{{C.5}{0}}
\@namedef{r@app:simulation_config}{{D}{0}}
\@namedef{r@app:sim_wireless}{{D.1}{0}}
\@namedef{r@app:sim_configurations}{{D.2}{0}}
\@namedef{r@app:sim_large}{{D.3}{0}}
\makeatother

\begin{document}

\title{Throughput-Optimal Control of Distributed Computing Networks for Service DAGs}

\author{Peng~Li,~\IEEEmembership{Graduate Student Member,~IEEE}, Jianan~Zhang,~\IEEEmembership{Member,~IEEE}, Jaime~Llorca,~\IEEEmembership{Senior Member,~IEEE}, Xinzhe~Fu, Xiang~Cheng,~\IEEEmembership{Fellow,~IEEE}, and Chen~Chen,~\IEEEmembership{Senior Member,~IEEE}%
\IEEEcompsocitemizethanks{\IEEEcompsocthanksitem Peng Li, Jianan Zhang, Xiang Cheng, and Chen Chen are with the State Key Laboratory of Photonics and Communications, School of Electronics,
Peking University, Beijing, China. Jaime Llorca is with the Centre Tecnol\`ogic de Telecomunicacions de Catalunya, Spain, and with New York University, USA. Xinzhe Fu is with Five Rings Capital, NY, USA.\protect\\
E-mail: zhangjianan@pku.edu.cn (corresponding author: Jianan Zhang).}%
}

\IEEEtitleabstractindextext{%
\begin{abstract}
Executing service workflows with directed acyclic graph (DAG) dependencies over wireless distributed computing networks requires joint control of function placement, routing, and scheduling, together with synchronization of inputs from the same request at merge points. We study throughput-optimal control of these networks under stochastic request arrivals, time-varying wireless links, and interference constraints. We use a valid mapping abstraction that assigns each service function to a computation node and each DAG dependency to a routing path. Using this abstraction, we characterize the network capacity region to support service DAGs, and prove that synchronizing the packets of a request at merge points does not shrink the region. We develop Dynamic Computing Network Control for service DAGs (DCNC-DAG), a dynamic control policy that selects queue-aware valid mappings and schedules service DAG execution by nearest-to-origin (NTO) while enforcing input synchronization at merge points. With exact minimum-cost valid mappings, DCNC-DAG is throughput-optimal and ensures rate stability of the physical queues under bounded arrivals. These guarantees hold for any fixed finite interval between mapping updates, even when mapping decisions rely on boundedly stale queue length information. For larger service DAGs, we develop DCNC-DAG-H, which replaces exact mapping optimization with a queue-aware heuristic. Simulations show that DCNC-DAG supports $2.7\times$--$5.5\times$ the throughput of queue-blind baselines, while DCNC-DAG-H achieves up to $2.68\times$ those of the corresponding baselines on larger workflow and AI service DAGs.
\end{abstract}

}

\maketitle
\IEEEdisplaynontitleabstractindextext
\IEEEpeerreviewmaketitle

\IEEEraisesectionheading{\section{Introduction}\label{sec:introduction}}

\IEEEPARstart{E}{merging} mobile and edge services increasingly execute workflows whose functions are coupled by data dependencies. Examples include embodied artificial intelligence (AI)~\cite{Embodied_AI_survey}, connected and autonomous driving~\cite{AD_survey}, and distributed mixture-of-experts (MoE) serving, which spreads experts across devices~\cite{MegaScaleInfer2025,LLEP2026}. A request may collect several raw inputs, execute interdependent functions, split intermediate results across branches, and later merge them. A directed acyclic graph (DAG) represents this structure, with service functions as vertices and dependency relations as edges. We refer to such a dependency graph as a \emph{service DAG}. When no single device can execute the complete request, both the functions and their dependency flow must be distributed across wireless access, backhaul, and edge computing resources.

We study a distributed computing network serving stochastic request arrivals over time-varying wireless links with interference constraints. For each service request, the controller decides execution nodes for service functions and routes for dependencies. Across requests, it schedules computation and wireless transmissions subject to the interference constraints. The objective is to support the largest set of request arrival rates---equivalently, to obtain throughput-optimal control for service DAGs.

In a general service DAG, different dependency edges may carry different data, and a downstream function can execute only after all required inputs of the same request have arrived. A capacity region characterization must therefore capture the coupling between function placement decisions and the routing decisions for different dependencies. The controller must also stabilize the synchronization backlog that accumulates at merge points when required inputs arrive at different times, in addition to the communication and computation queues.

Existing throughput-optimal control algorithms address service function chains in cloud networks~\cite{DCNC2016,DCNC_journal2018}, mixed-cast service flows~\cite{UCNC}, generalized network flow problems~\cite{UMW}, and DAG jobs with precedence constraints on multiple servers~\cite{Pedarsani2014DAG}. Synchronization models typically assume fixed placement and routing~\cite{DaiLin2005MaximumPressure,Baccelli1989ForkJoin}. Neither line of work jointly addresses function placement, per-dependency routing, wireless scheduling under interference constraints, and request synchronization in the service DAG setting.

We address this gap through a valid mapping~\cite{DynVMP}, which assigns each service function to a computation node and each dependency to a network path. The resulting communication and computation load vectors provide the basis for a state-averaged linear programming (LP) characterization of the capacity region. Our Dynamic Computing Network Control for service DAGs (DCNC-DAG) selects mappings using virtual queue lengths as congestion prices and activates wireless links by a max-weight rule. The selected mappings are executed in the physical system using nearest-to-origin (NTO) scheduling~\cite{NTO_origin} for transmissions and computations, together with per-request waiting pools that synchronize inputs at merge points. Since exact minimum-cost valid mapping is NP-hard for general service DAGs~\cite{VNEP_hardness}, we also develop DCNC-DAG-H, a queue-aware heuristic for larger service DAGs.

The main contributions of this paper are summarized as follows.
\begin{itemize}
    \item We characterize the capacity region of distributed computing networks serving arbitrary service DAGs under i.i.d.\ wireless states and interference constraints on link activation. By decomposing each completed request into valid mappings that preserve its function placements, we derive a state-averaged LP whose interior remains achievable despite request synchronization at merge points.

    \item We develop DCNC-DAG, a frame-based dynamic control policy that makes queue-aware function placement and routing decisions. It schedules wireless transmissions and computations while synchronizing inputs of the same request at merge points. We further develop DCNC-DAG-H, a queue-aware heuristic variant for larger service DAGs.

    \item For any finite frame length and bounded i.i.d.\ request arrivals, we prove that DCNC-DAG with exact minimum-cost valid mappings is throughput-optimal: the virtual queues are strongly stable, while the physical queues and synchronization waiting pools are rate stable. A key step in the physical stability proof is to bound the number of packets waiting at merge points. The same guarantees hold with boundedly stale queue prices.

    \item DCNC-DAG is evaluated on four wireless networks and achieves $2.7\times$--$5.5\times$ the throughput of queue-blind baselines, while DCNC-DAG-H is evaluated on workflow and AI service DAGs with 20--31 functions and achieves up to $2.68\times$ the throughput of the baselines in the corresponding experiments.
\end{itemize}

\section{Related Work}\label{sec:related_work}

\emph{Throughput-optimal network control.}
Dynamic cloud network control (DCNC)~\cite{DCNC2016,DCNC_journal2018} establishes throughput-optimal control for service chains, UCNC~\cite{UCNC} extends joint placement and routing to mixed unicast and multicast traffic, and UMW~\cite{UMW} accommodates time-varying links and general feasible activation sets. Their path, tree, or generalized flow representations do not capture service DAGs requiring inputs from multiple upstream functions to be synchronized within each request. Extensions address software-defined networking and network function virtualization integration~\cite{TCC2023DynamicComputation}, reconfiguration costs~\cite{ReconfigDelayCost2019}, and fixed data-intensive pipelines with multi-hop routing and identity queues~\cite{Joint3CServiceDelivery}, but retain this structural limitation. Throughput-optimal scheduling for DAG jobs~\cite{Pedarsani2014DAG} handles precedence and synchronization on multiple servers without multi-hop routing or wireless dynamics. Dispersed computing~\cite{Yang2019Dispersed} extends scheduling of serial tasks to DAGs with input joins under broadcast communication, but does not address per-dependency multi-hop routing or wireless link activation under interference constraints.

\emph{Synchronization with fixed placement and routing.}
Maximum-pressure control for stochastic processing networks~\cite{DaiLin2005MaximumPressure} and stability models for acyclic fork-join networks~\cite{Baccelli1989ForkJoin} account for processing dependencies and merge synchronization under fixed task locations and routing relations. They do not jointly choose function placement and per-dependency routing or handle transmissions coupled by wireless interference. Placement and routing decisions determine when the inputs of each request arrive at a merge point, and the resulting waiting backlog must also be stabilized.

\emph{DAG placement and wireless computing.}
Workflow scheduling~\cite{WorkflowWirelessSmallCloud2018}, cooperative DAG offloading~\cite{DAGSoftCoop2024}, and mobility-aware offloading~\cite{MobilityAwareDepOffload2025} primarily use offline or one-shot formulations. DAG orchestration over the compute continuum~\cite{IDAGO} and Jupiter~\cite{JupiterTP} address DAG execution without throughput-optimal online control under stochastic arrivals and network conditions. Virtual network embedding~\cite{VNEP_hardness,DynVMP} provides offline placement and routing methods; we use DynVMP within our online controller. Learning-based device placement~\cite{GiPH} focuses on placement. Wireless computing control~\cite{WirelessComputing2018}, dense cell offloading~\cite{MECOffloadDense2018}, and dynamic service provisioning~\cite{DynamicServiceProvisioning2023} incorporate wireless or resource dynamics, but their guarantees do not cover the joint placement, per-dependency routing, and input synchronization considered here.

\emph{Delayed and infrequent control information.}
Throughput can be preserved under delayed network state~\cite{YingDelayedNSI2011}, frame-based decisions~\cite{NeelyDelayComplexity2002,NeelyRenewal2013}, and reconfiguration delay~\cite{WangJavidiReconfigDelay2017}. Most closely related is~\cite{GopalanDelayedQueue2013}, which separates slowly updated queue information from fast channel scheduling. In DCNC-DAG, the slow decision is a valid mapping, so the analysis must bound the effect of stale prices across all communication and computation resources it uses.

\begin{table*}[htbp]
\centering
\caption{Comparison of closely related works.}
\label{tab:related_compare}
\footnotesize
\setlength{\tabcolsep}{4pt}
\renewcommand{\arraystretch}{1.1}
\begin{tabularx}{\textwidth}{@{}l LLLLL@{}}
\toprule
Reference
& Service structure
& Multi-hop routing
& Wireless \& interference
& Input matching
& Stability / throughput guarantee \\
\midrule

DCNC~\cite{DCNC2016,DCNC_journal2018}
& chain / tree
& yes
& no
& by type
& throughput-optimal \\

UCNC~\cite{UCNC}
& chain
& yes
& no
& none
& throughput-optimal \\

UMW~\cite{UMW}
& --
& yes
& yes
& none
& throughput-optimal \\

Processing networks~\cite{DaiLin2005MaximumPressure}
& processing network
& no
& no
& by type
& throughput-optimal \\

Fork-join networks~\cite{Baccelli1989ForkJoin}
& DAG
& fixed
& no
& by request
& stability analysis \\

DAG job scheduling~\cite{Pedarsani2014DAG}
& DAG
& no
& no
& by request
& throughput-optimal \\

Dispersed computing~\cite{Yang2019Dispersed}
& chain / DAG
& no
& no
& by request
& throughput-optimal \\

Data-intensive pipelines~\cite{Joint3CServiceDelivery}
& chain / tree
& yes
& no
& by request
& throughput-optimal \\

\textbf{This paper}
& \textbf{DAG}
& \textbf{yes}
& \textbf{yes}
& \textbf{by request}
& \textbf{throughput-optimal} \\

\bottomrule
\end{tabularx}

\tabnote{``fixed'' denotes predetermined routes.
Input matching is by input type or by request identity,
including job identity; ``none'' denotes no multi-input matching.
``--'' indicates that computation is not modeled.}
\end{table*}

To the best of our knowledge, DCNC-DAG is the first policy with a throughput optimality guarantee for service DAGs that jointly handles function placement, per-dependency multi-hop routing, per-request input synchronization, and wireless transmission under time-varying link states and interference constraints. Table~\ref{tab:related_compare} summarizes the differences from closely related work.

\section{System Model}\label{sec:system_model}

In this section, we introduce the network, service DAG, and traffic models and define the dynamic control problem.

\subsection{Distributed Computing Network Model}\label{subsec:computing_network_model}

We model the distributed computing network as a directed graph \(\mathcal{G}=(\mathcal{V},\mathcal{E})\), where \(\mathcal{V}\) is the node set and \(\mathcal{E}\) is the set of directed wireless links. Each node can forward packets and host one or more service functions. Time is slotted with slots indexed by \(t\in\{0,1,2,\ldots\}\). Let \(\mu_u\) denote the computation capacity of node \(u\in\mathcal V\), and let \(\mu_{uv}\) denote the communication capacity of link \((u,v)\in\mathcal E\) when it is available. Let \(\sigma_{uv}(t)\in\{0,1\}\) denote the wireless state of link \((u,v)\) in slot \(t\), where \(\sigma_{uv}(t)=1\) indicates the link is ON and \(\sigma_{uv}(t)=0\) indicates the link is OFF. The instantaneous communication capacity is \(\mu_{uv}(t)=\mu_{uv}\sigma_{uv}(t)\). Let \(\boldsymbol{\sigma}(t)=\{\sigma_{uv}(t),(u,v)\in\mathcal{E}\}\) denote the wireless state vector. For a network node \(u\), let \(\delta^{+}(u)\) and \(\delta^{-}(u)\) denote its outgoing and incoming neighbors in \(\mathcal G\), respectively.

To model wireless interference, let \(\ActSet \subseteq \{0,1\}^{|\mathcal{E}|}\) be the set of feasible link activation vectors. For \(\mathbf{s}(t)\in\ActSet\), \(s_{uv}(t)=1\) indicates that link \((u,v)\) is activated in slot \(t\). Combining the observed ON/OFF states with the interference constraints, the feasible activation set is
\begin{equation}
\begin{aligned}
    \ActSet(\boldsymbol{\sigma}(t))
    \triangleq
    \left\{
    \mathbf{s}\in \ActSet:
    s_{uv}\le \sigma_{uv}(t),\ \forall (u,v)\in\mathcal{E}
    \right\}.
    \nonumber
\end{aligned}
\end{equation}

If \(\mathbf{s}(t)\in\ActSet(\boldsymbol{\sigma}(t))\) is selected in slot \(t\), link \((u,v)\) can serve up to \(\mu_{uv}(t)s_{uv}(t)\) workload units in slot \(t\).

\begin{assumption}\label{assumption:wireless_state}
The wireless state vectors \(\{\boldsymbol{\sigma}(t)\}_{t\geq 0}\) are i.i.d.\ over time.
\end{assumption}

\begin{remark}\label{remark:primary}
The feasible activation set \(\ActSet\) accommodates general link interference constraints. Specifically, under the primary interference model, each node can be incident to at most one activated link in a slot:
\begin{equation}
\begin{aligned}
    \ActSet_{\rm pri}
    \!=\!\!
    \left\{\!
    \mathbf{s}\!\in\!\{0,1\}^{|\mathcal{E}|}\!:\!\!
    \sum_{v\in\delta^{+}(u)}\!s_{uv}
    \!+\!\!
    \sum_{v\in\delta^{-}(u)}\!s_{vu}
    \!\le\! 1,
    \forall u\!\in\!\mathcal{V}\!
    \right\}.
    \nonumber
\end{aligned}
\end{equation} 
\end{remark}

\subsection{Service DAG Model}\label{subsec:service_DAG_model}

A service DAG is represented by a directed acyclic graph \(\phi=(\mathcal V^\phi,\mathcal E^\phi)\), where \(\mathcal V^\phi\) contains service functions and \(\mathcal E^\phi\) contains dependency relations. A directed edge \((A^\phi,B^\phi)\) indicates that processing \(B^\phi\) requires the output produced by \(A^\phi\). For each service function \(A^\phi\), let \(\mathcal V_{A^\phi}\subseteq\mathcal V\) denote its qualified node set, i.e., the computation nodes that host an implementation of \(A^\phi\) and hence can execute it. We use the same notation \(\delta^{+}(A^\phi)\) and \(\delta^{-}(A^\phi)\) for the successor and predecessor functions of \(A^\phi\) in \(\phi\), respectively.

Source functions have no predecessor functions in the service DAG and take raw inputs, whereas sink functions produce outputs that leave the network. Let \(\mathcal V_s^\phi\) and \(\mathcal V_d^\phi\) denote the sets of source and sink functions, respectively. For each source function \(A^\phi\in\mathcal V_s^\phi\), introduce an auxiliary service source \(S_A^\phi\); similarly, for each sink function \(A^\phi\in\mathcal V_d^\phi\), introduce an auxiliary service sink \(D_A^\phi\). Define
\[
\mathcal S^\phi=\{S_A^\phi:A^\phi\in\mathcal V_s^\phi\},\qquad \mathcal D^\phi=\{D_A^\phi:A^\phi\in\mathcal V_d^\phi\},
\]
with
\[
\mathcal E_s^\phi=\{(S_A^\phi,A^\phi):A^\phi\in\mathcal V_s^\phi\},\enspace \mathcal E_d^\phi=\{(A^\phi,D_A^\phi):A^\phi\in\mathcal V_d^\phi\}.
\]
The augmented service DAG has node set \(\bar{\mathcal V}^\phi=\mathcal V^\phi\cup\mathcal S^\phi\cup\mathcal D^\phi\) and edge set \(\bar{\mathcal E}^\phi=\mathcal E^\phi\cup\mathcal E_s^\phi\cup\mathcal E_d^\phi\).

For each augmented dependency edge \((A^\phi,B^\phi)\in\bar{\mathcal E}^\phi\), let \(\xi^{(A^\phi,B^\phi)}>0\) denote the communication workload carried from \(A^\phi\) to \(B^\phi\) per request. For each service function \(A^\phi\in\mathcal V^\phi\), let \(r^{A^\phi}>0\) denote the computation workload required to execute the function. For each request, define \(\boldsymbol\xi_s^\phi=\{\xi^{(S_A^\phi,A^\phi)}:A^\phi\in\mathcal V_s^\phi\}\) and \(\boldsymbol\xi_d^\phi=\{\xi^{(A^\phi,D_A^\phi)}:A^\phi\in\mathcal V_d^\phi\}\) as the input workloads required by the source functions and the output workloads delivered by the sink functions, respectively.

For modeling simplicity, each request is an atomic unit: every service function is executed exactly once at one qualified node, without splitting that execution across nodes. A flow comprises multiple such requests, which may assign the same service function to different qualified nodes and execute concurrently. Execution starts only after all of its required inputs have arrived at that node. After execution, the function produces its outputs once and sends them toward its successor functions. Merging and execution consume at most one copy of each required packet, so duplicate packets cannot create additional usable outputs. Data associated with different dependency edges are distinct and routed separately; in particular, outputs for different successor functions differ in content and therefore cannot benefit from multicast.

\begin{figure}[htbp]
    \centering
    \subfloat[Dependency structure.]{\label{subfig:service_DAG_a}
        \includegraphics[width=0.95\linewidth]{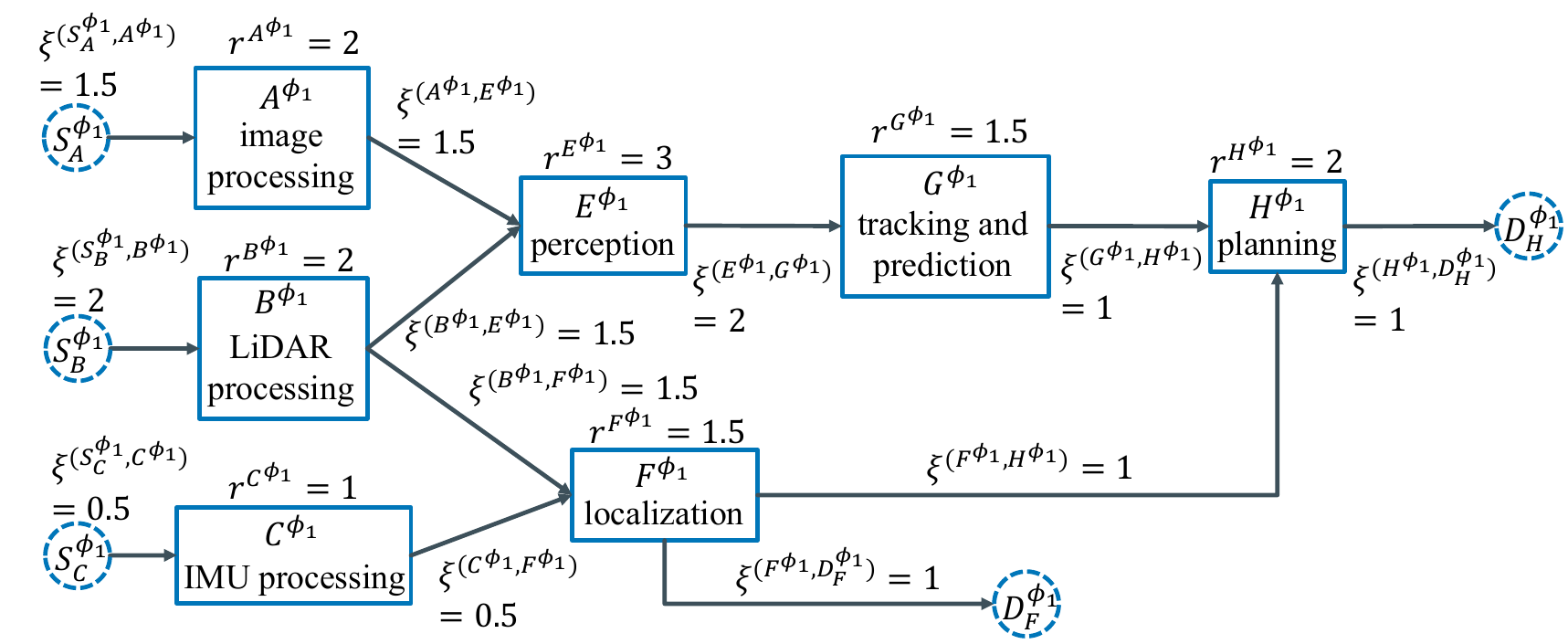}}\\
    \subfloat[Merging, splitting, and workload scaling.]{\label{subfig:service_DAG_b}
        \includegraphics[width=0.7\linewidth]{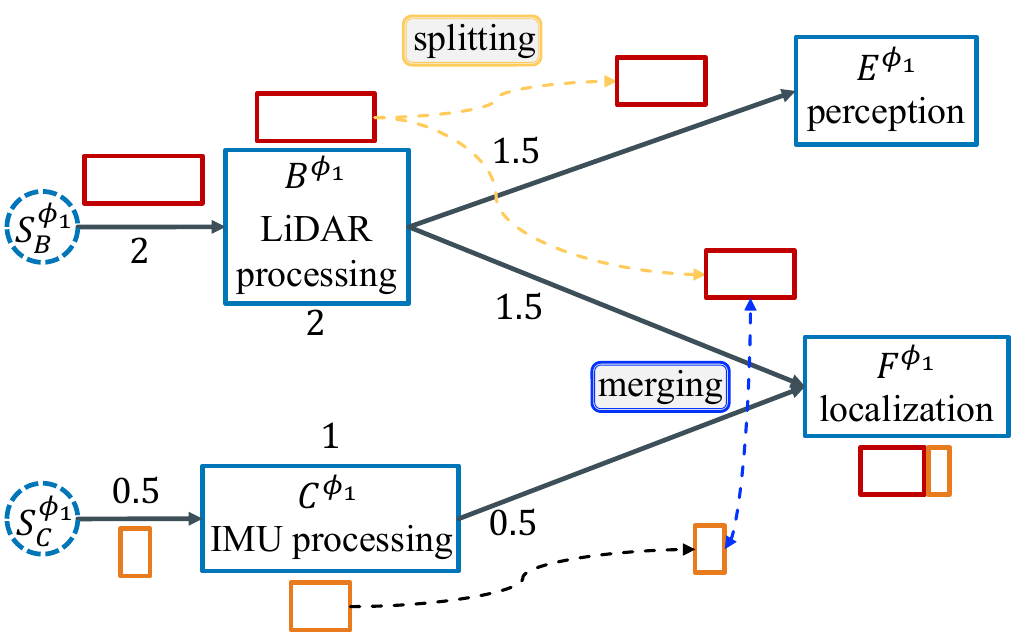}}
    \caption{Service DAG of an autonomous driving task. A request merges multiple raw inputs, splits intermediate outputs across successor functions, and may change the workload along different dependencies; the numerical values are illustrative.}
    \label{fig:service_DAG}
\end{figure}

Fig.~\ref{fig:service_DAG} illustrates the model through an autonomous driving task with image, LiDAR, and IMU inputs, perception, localization, tracking, and planning functions, and control command outputs. The example illustrates input merging, output splitting, and workload scaling across dependencies.

\subsection{Traffic Model}\label{subsec:traffic_model}

A commodity \((c,\phi)\) is specified by a service DAG \(\phi\) and its endpoints in the computing network: \(\rho_s^{(c,\phi)}:\mathcal S^{\phi}\to\mathcal V\) and \(\rho_d^{(c,\phi)}:\mathcal D^{\phi}\to\mathcal V\). Let \(\mathcal K\) denote the finite commodity set, and write \(S_A^c\triangleq \rho_s^{(c,\phi)}(S_A^{\phi})\), \(D_A^c\triangleq \rho_d^{(c,\phi)}(D_A^{\phi})\). A request is completed once all sink outputs have been fully delivered to their corresponding network endpoints.

Let \(A^{(c,\phi)}(t)\) denote the number of commodity-\((c,\phi)\) requests arriving in slot \(t\). A request may inject source packets at multiple service sources.

\begin{assumption}[Proportional request arrivals]\label{assumption:proportional_arrivals}
Let \(\mathbf A(t)\triangleq\bigl(A^{(c,\phi)}(t)\bigr)_{(c,\phi)\in\mathcal K}\). The process \(\{\mathbf A(t)\}_{t\ge0}\) is i.i.d.\ across slots and independent of the wireless state process \(\{\boldsymbol\sigma(t)\}_{t\ge0}\). For each commodity, \(\mathbb E[A^{(c,\phi)}(t)]=\lambda^{(c,\phi)}\), and the workload vector contributed by its source functions in slot \(t\) is \(A^{(c,\phi)}(t)\boldsymbol\xi_s^\phi\), with mean \(\lambda^{(c,\phi)}\boldsymbol\xi_s^\phi\).
\end{assumption}

This model assumes that all source packets of a request are generated in the same slot, as when sensors are triggered by a common hardware clock. The assumption can be relaxed to allow for uniformly bounded differences in the arrival times of source packets belonging to the same request, and inputs belonging to the same request can be merged. Under asynchronous arrivals, the source packets belonging to the same request are held until all inputs arrive and are then released together into the network. Under bounded request arrivals, the number of requests awaiting release remains uniformly bounded, preserving the same stability and throughput optimality guarantees.

The vector $\boldsymbol\xi_s^\phi$ specifies the number of source packets required by one request. Unequal raw input rates can be accommodated by forming requests from the required source inputs and discarding surplus inputs that cannot contribute to additional request completions.
\begin{assumption}[Bounded request arrivals]\label{assumption:fixedK_bounded_arrival}
For each commodity \((c,\phi)\), there exists a finite constant \(A_{\max}^{(c,\phi)}\) such that \(A^{(c,\phi)}(t)\le A_{\max}^{(c,\phi)}\) for all \(t\), almost surely.
\end{assumption}

\begin{definition}[Dynamic control problem for service DAGs]
Consider a network \(\mathcal{G}\) with node capacities \(\{\mu_u\}\), link capacities \(\{\mu_{uv}\}\), wireless state process \(\{\boldsymbol\sigma(t)\}\), feasible activation set \(\ActSet\), and commodity set \(\mathcal K\). The dynamic control problem for service DAGs is to choose processing locations, dependency routes, link activations, and schedules in each slot. The objective is to support the largest set of request arrival rates, subject to the service DAG dependencies, the wireless constraints \(\mathbf{s}(t)\in\ActSet(\boldsymbol\sigma(t))\), the link capacities \(\mu_{uv}(t)s_{uv}(t)\), and the node capacities \(\mu_u\).
\end{definition}

The controller is logically centralized. In each slot, it observes the request arrivals \(\mathbf A(t)\) and wireless state \(\boldsymbol\sigma(t)\), and maintains virtual queue lengths as congestion prices from these observations and its previous decisions. Mapping decisions use the prices observed at the beginning of each frame, while link activation uses the current prices and wireless state every slot. Sections~\ref{sec:routing_policy} and~\ref{sec:physical_network} specify these rules and their physical execution.

\section{Capacity Region}\label{sec:capacity_region}

In this section, we use the valid mapping of \cite{DynVMP} to represent the placement and routing decisions of one request and characterize the network capacity region by a state-averaged LP under wireless interference and time-varying link states.

\subsection{Valid Mapping}\label{subsec:valid_mapping}

\begin{definition}[Valid mapping \cite{DynVMP}]
A valid mapping of commodity \((c,\phi)\) is a tuple \(m^{(c,\phi)}=(m^{(c,\phi)}_{\mathcal{V}},m^{(c,\phi)}_{\mathcal{E}})\), where \(m^{(c,\phi)}_{\mathcal{V}}\) maps nodes of the augmented service DAG to computation nodes in \(\mathcal G\) and \(m^{(c,\phi)}_{\mathcal{E}}\) maps dependency edges to simple paths in \(\mathcal G\). The qualified node sets \(\mathcal V_{A^{\phi}}\) are those of Section~\ref{subsec:service_DAG_model}; auxiliary sources and sinks have singleton candidate sets \(\{S_A^c\}\) and \(\{D_A^c\}\). The mapping must satisfy:
\begin{itemize}
    \item \emph{Node mapping}: for each $A^{\phi}\in\bar{\mathcal{V}}^{\phi}$, $m^{(c,\phi)}_{\mathcal{V}}(A^{\phi})\in\mathcal{V}_{A^{\phi}}$;
    \item \emph{Edge mapping}: for each \((A^{\phi},B^{\phi})\in\bar{\mathcal{E}}^{\phi}\) such that \(m_{\mathcal V}^{(c,\phi)}(A^{\phi})\neq m_{\mathcal V}^{(c,\phi)}(B^{\phi})\), \(m_{\mathcal E}^{(c,\phi)}(A^{\phi},B^{\phi})\) is a simple path in \(\mathcal G\) from \(m_{\mathcal V}^{(c,\phi)}(A^{\phi})\) to \(m_{\mathcal V}^{(c,\phi)}(B^{\phi})\); co-located dependencies map to the empty path and consume no link capacity.
\end{itemize}
\end{definition}

Let \(\mathcal{M}^{(c,\phi)}\) denote the finite set of valid mappings of commodity \((c,\phi)\). We retain only mappings that use available resources, i.e., computation nodes \(u\) with \(\mu_u>0\) and links that can be activated with positive probability under some feasible wireless state and activation. Removing all other mappings does not affect the capacity region. We assume that the resulting set \(\mathcal{M}^{(c,\phi)}\) is nonempty.

For a valid mapping \(m_{k}^{(c,\phi)}\), define the corresponding per-request load vector by
\begin{equation}
\begin{aligned}
    \mathtt{M}^{(c,\phi)}_{k,uv}
    &=
    \sum_{(A^\phi,B^\phi)\in\bar{\mathcal E}^{\phi}}
    \xi^{(A^{\phi},B^{\phi})}
    \mathbb{I}\left\{
    (u,v)\in m_{k,\mathcal{E}}^{(c,\phi)}(A^{\phi},B^{\phi})
    \right\},
    \\
    \mathtt{M}^{(c,\phi)}_{k,u}
    &=
    \sum_{A^\phi\in\mathcal V^{\phi}}
    r^{A^{\phi}}
    \mathbb{I}\{u=m_{k,\mathcal{V}}^{(c,\phi)}(A^{\phi})\}.
    \nonumber
\end{aligned}
\end{equation}

A valid mapping specifies the placement and routing decisions for one request. It is the per-request control action used by DCNC-DAG, and its load vector is the basic accounting unit of the capacity region characterization developed below.

\subsection{Policy Space}\label{subsec:capacity_region}

To connect packet-level routing with the communication workload model, we use a normalized packet unit. We assume that the communication workloads are rational numbers; since the commodity set is finite, the unit can be chosen so that every augmented dependency edge carries an integer number of normalized packets per request. We measure communication workloads in normalized packet units and link capacities in the corresponding units per slot. Thus, transmitting one normalized packet over one link requires one unit of transmission service.

\begin{definition}[Admissible policy]
An admissible dynamic control policy $\pi$ controls two actions:
(i) \emph{routing and processing}: for each request, choose one qualified execution node for every service function and route the packets associated with every dependency edge;
(ii) \emph{scheduling}: in every slot $t$, choose \(\mathbf{s}(t)\in\ActSet(\boldsymbol{\sigma}(t))\) and schedule transmissions and processing under capacities \(\mu_{uv}(t)s_{uv}(t)\) and \(\mu_{u}\). 
\end{definition}


Let \(\Pi\) denote the set of admissible policies and, for \(\pi\in\Pi\), let \(\kappa^{(c,\phi),\pi}(t)\) denote the number of commodity-\((c,\phi)\) requests completed by time \(t\) under policy \(\pi\), where completion is determined for each request according to the rule in Section~\ref{subsec:traffic_model}. Outputs belonging to different requests cannot be combined to form a completion. A policy \(\pi\in\Pi\) is said to support an arrival rate vector \(\boldsymbol{\Lambda}=\{\lambda^{(c,\phi)}\}\) if \(\liminf_{t\to\infty}\kappa^{(c,\phi),\pi}(t)/t=\lambda^{(c,\phi)}\) w.p.\,1 for every \((c,\phi)\in\mathcal K\).

\begin{definition}[Capacity region]
The capacity region \(\Gamma(\mathcal{G},\mathcal K)\) is the set of arrival rate vectors \(\boldsymbol{\Lambda}\) supported by some admissible policy \(\pi\in\Pi\).
\end{definition}

A policy is \emph{throughput-optimal} if it supports every arrival rate vector in \(\mathrm{int}(\Gamma(\mathcal G,\mathcal K))\).

Under interference, the long-term communication capacities of different links are coupled. For each wireless state \(\sigma\), let \(p_\sigma=\Pr\{\boldsymbol{\sigma}(t)=\sigma\}\). For each \(s\in\ActSet(\sigma)\), let \(\alpha_{\sigma,s}\ge0\) denote the fraction of slots with state \(\sigma\) in which activation \(s\) is selected. The state-averaged communication service region is
\[
\begin{aligned}
\overline{\mathcal C}=\Bigl\{\bar{\mathbf c}\in\mathbb R_+^{|\mathcal E|}:{}&
\exists\,\alpha_{\sigma,s}\ge0,\quad
\sum_{s\in\ActSet(\sigma)}\alpha_{\sigma,s}\le1,\ \forall\sigma,\\
&\bar c_{uv}\le
\sum_{\sigma,\,s\in\ActSet(\sigma)}
p_\sigma\alpha_{\sigma,s}\mu_{uv}\sigma_{uv}s_{uv},\\[-1mm]
&\hspace{25mm}\forall(u,v)\in\mathcal E
\Bigr\}.
\end{aligned}
\]

\subsection{Capacity Region}\label{subsec:lp_characterization}

The admissible policy space is more general than policies that explicitly select valid mappings. The following lemma connects arbitrary completed executions to the valid mapping load representation used in the capacity characterization.

\begin{lemma}[Valid mapping decomposition of a completed request]\label{lemma:decomposition_main}
For every completed request, there exists a convex combination of valid mappings that all place each service function at the node where it was executed for that request. The communication load induced by this combination on each link does not exceed the load actually incurred by the request on that link, whereas the induced computation load on each node equals the request's actual computation load on that node.
\end{lemma}

The construction extends the efficient route reduction used in UCNC~\cite{UCNC}. For a service DAG, the path decompositions of different dependencies must additionally remain consistent with the same function placements of a request; Lemma~\ref{lemma:decomposition_main} preserves this consistency while keeping the computation loads unchanged and not increasing the communication loads. The complete proof is given in Appendix~\ref{app:capacity}.

Let \(\Gamma_{\rm LP}\) be the set of arrival rate vectors \(\boldsymbol{\Lambda}\) for which there exist rates \(\{f_{k}^{(c,\phi)}\ge0\}\) of valid mappings and \(\bar{\mathbf c}\in\overline{\mathcal{C}}\) with
\begin{equation}
\begin{aligned}
    \lambda^{(c,\phi)}
    =
    \sum_{k}f_{k}^{(c,\phi)},
    \quad
    \forall (c,\phi)\in\mathcal K,
    \label{eq:flow_decomposition}
\end{aligned}
\end{equation}
\begin{equation}
\begin{aligned}
    \sum_{(c,\phi),k}
    f^{(c,\phi)}_k\,\mathtt{M}_{k,uv}^{(c,\phi)}
    \leq
    \bar{c}_{uv},
    \quad
    \forall (u,v)\in\mathcal{E},
\end{aligned}
\end{equation}
\begin{equation}
\begin{aligned}
    \sum_{(c,\phi),k}
    f^{(c,\phi)}_{k}\,\mathtt{M}_{k,u}^{(c,\phi)}
    \leq
    \mu_{u},
    \quad
    \forall u\in\mathcal{V}.
    \label{eq:node_capacity_constraint}
\end{aligned}
\end{equation}

The next theorem identifies the resource constraints that limit achievable throughput. Every supportable arrival rate must satisfy the LP, and every rate strictly inside its feasible region is achievable under bounded arrivals, despite synchronization at merge points.

\begin{theorem}[Capacity region characterization]\label{theorem:capacity}
Under Assumptions~\ref{assumption:wireless_state}--\ref{assumption:proportional_arrivals}, every arrival rate vector supportable by an admissible policy belongs to \(\Gamma_{\rm LP}\). If Assumption~\ref{assumption:fixedK_bounded_arrival} also holds, every vector in \(\mathrm{int}(\Gamma_{\rm LP})\) is supportable. Consequently, \(\Gamma_{\rm LP}\) characterizes the capacity region \(\Gamma(\mathcal G,\mathcal K)\) up to a possible difference on their boundaries, and
\[
    \mathrm{int}(\Gamma(\mathcal G,\mathcal K))
    =
    \mathrm{int}(\Gamma_{\rm LP}).
\]
\end{theorem}

Appendix~\ref{app:capacity} proves necessity by applying Lemma~\ref{lemma:decomposition_main} to completed requests and taking limits of the empirical mapping rates. The constructive sufficiency direction follows from the virtual queue and physical queue stability results below and Corollary~\ref{cor:throughput}.

The LP serves as an analytical characterization of the capacity region. The online policy developed next does not solve this LP and requires no knowledge of the arrival rates \(\{\lambda^{(c,\phi)}\}\) or the wireless state distribution \(\{p_\sigma\}\).

\section{Dynamic Routing Policy for a Virtual System}\label{sec:routing_policy}

This section introduces the virtual queues and frame-based mapping policy, establishes virtual queue stability under current or boundedly stale mapping prices, and describes a heuristic for larger service DAGs.

\subsection{Virtual Queueing System}\label{subsec:virtual_system}

Virtual queues are auxiliary workload counters used for control and stability analysis; they are distinct from the actual system queues. Each link \((u,v)\) has a virtual communication queue \(\widetilde{Q}_{uv}\), and each node \(u\) a virtual computation queue \(\widetilde{Q}_{u}\). When a request arrives and a valid mapping is selected, the virtual system immediately charges the full communication and computation workloads induced by that mapping to the corresponding virtual queues, without accounting for the sequential order of the physical transmissions and computations. Let \(k^{(c,\phi),\pi}(t)\) denote the index of the mapping selected by policy \(\pi\) for commodity \((c,\phi)\) in slot \(t\). The resulting virtual arrivals are
\(\widetilde{A}_{uv}^{(c,\phi),\pi}(t)=A^{(c,\phi)}(t)\mathtt{M}^{(c,\phi)}_{k^{(c,\phi),\pi}(t),uv}\) and \(\widetilde{A}_{u}^{(c,\phi),\pi}(t)=A^{(c,\phi)}(t)\mathtt{M}^{(c,\phi)}_{k^{(c,\phi),\pi}(t),u}\), summed over commodities to give \(\widetilde{A}_{uv}^{\pi}(t)\), \(\widetilde{A}_{u}^{\pi}(t)\). The virtual queues evolve as
\begin{equation}
\begin{aligned}
    \widetilde{Q}_{uv}(t+1)
    =
    \left(
    \widetilde{Q}_{uv}(t)
    +
    \widetilde{A}_{uv}^{\pi}(t)
    -
    \mu_{uv}(t)s_{uv}^{\pi}(t)
    \right)^{+},
    \label{eq:virtual_Quv_evolution}
\end{aligned}
\end{equation}
\begin{equation}
\begin{aligned}
    \widetilde{Q}_{u}(t+1)
    =
    \left(
    \widetilde{Q}_{u}(t)
    +
    \widetilde{A}_{u}^{\pi}(t)
    -
    \mu_{u}
    \right)^{+},
    \label{eq:virtual_Qu_evolution}
\end{aligned}
\end{equation}
where \((a)^{+}=\max(a,0)\) and \(\mathbf s^{\pi}(t)\in\ActSet(\boldsymbol\sigma(t))\).

\subsection{Frame-Based Routing Policy}\label{subsec:fixedK_policy}

Fix a positive integer \(\FrameLen\) and partition time into frames \(\mathcal T_n=\{n\FrameLen,\ldots,n\FrameLen+\FrameLen-1\}\). Let \(\framehead=n\FrameLen\) and \(\Qhat(n)=\widetilde Q(\framehead)\) be the virtual queue lengths observed at the beginning of frame \(n\) and used for mapping. Virtual queues continue to evolve within each frame, and link activation uses the current queue prices and wireless state every slot.

\begin{definition}[Routing policy \(\policyvk\)]
At the beginning of frame \(n\), the policy initializes an empty mapping table. In each slot \(t\in\mathcal{T}_{n}\):
(i) for each commodity \((c,\phi)\) with \(A^{(c,\phi)}(t)>0\) and no stored mapping in frame \(n\), compute
\begin{equation}
\begin{aligned}
    \widehat{k}_{n}^{(c,\phi)}
    \in{}&
    \arg\min_{k:\, m_{k}^{(c,\phi)}\in\mathcal{M}^{(c,\phi)}}
    \Biggl[
    \sum_{(u,v)\in\mathcal{E}}
    \Qhat_{uv}(n)\mathtt{M}^{(c,\phi)}_{k,uv}
    \\
    &\qquad\qquad\qquad\qquad
    +
    \sum_{u\in\mathcal{V}}
    \Qhat_{u}(n)\mathtt{M}^{(c,\phi)}_{k,u}
    \Biggr],
\end{aligned}
\label{eq:fixedK_mapping}
\end{equation}
store it, and assign all commodity-\((c,\phi)\) arrivals within the frame to it;
(ii) choose the link activation by the max-weight rule
\begin{equation}
\begin{aligned}
    \mathbf{s}^{*}(t)
    \in
    \arg\max_{\mathbf{s}\in\ActSet(\boldsymbol{\sigma}(t))}
    \sum_{(u,v)\in\mathcal{E}}
    \widetilde{Q}_{uv}(t)\mu_{uv}(t)s_{uv}.
    \label{eq:max_weight_activation}
\end{aligned}
\end{equation}
Under the primary interference model of Remark~\ref{remark:primary}, \eqref{eq:max_weight_activation} reduces to a maximum-weight matching over the currently ON links.
\end{definition}

When \(\FrameLen=1\), \(\Qhat(n)\) equals the current virtual queue vector, so \eqref{eq:fixedK_mapping} reduces to the minimum-cost valid mapping under the current virtual queues. To solve \eqref{eq:fixedK_mapping} exactly, we apply DynVMP~\cite{DynVMP} to the augmented service DAG. Each function uses its qualified node set as the candidate set, while each auxiliary source or sink has a singleton candidate set. For dependency \((A^{\phi},B^{\phi})\), physical link \((u,v)\) is assigned weight \(\xi^{(A^{\phi},B^{\phi})}\Qhat_{uv}(n)\), and placing \(A^{\phi}\) at node \(u\) incurs weight \(r^{A^{\phi}}\Qhat_u(n)\). A dependency whose endpoints are placed at the same node uses the empty path with zero cost. DynVMP solves this problem in \(O(|\mathcal{V}^{\phi}|^{3}|\mathcal{V}|^{2\tau+3})\) time~\cite{DynVMP}, where \(\tau\) is the width of a tree decomposition for the undirected graph underlying the augmented service DAG. Thus, exact mapping is polynomial when the treewidth is bounded by a constant.

For DCNC-DAG, using a frame length \(\FrameLen>1\) further amortizes the mapping computation cost by reusing each computed mapping for subsequent arrivals of the same commodity within the frame. However, minimum-cost valid mapping remains NP-hard for general service DAGs even without capacity constraints~\cite{VNEP_hardness}. DCNC-DAG-H reduces the cost of each mapping computation by replacing the exact solver with a queue-aware heuristic, as described in Section~\ref{subsec:heuristic_extension}.

\subsection{Virtual Queue Stability}\label{subsec:virtual_stability}

As virtual queues evolve within a frame, the stored mapping may cease to minimize the current queue-weighted cost.

For commodity \((c,\phi)\), define the cost of valid mapping \(k\) under the current virtual queues as
\[
\Psi_t^{(c,\phi)}(k)
\triangleq
\sum_{(u,v)\in\mathcal E}
\widetilde Q_{uv}(t)\mathtt M_{k,uv}^{(c,\phi)}
+
\sum_{u\in\mathcal V}
\widetilde Q_u(t)\mathtt M_{k,u}^{(c,\phi)}.
\]

For \(\FrameLen>1\), the following proposition holds under Assumption~\ref{assumption:fixedK_bounded_arrival}; for \(\FrameLen=1\), the mapping uses current queue prices and no such assumption is required. The proposition bounds the resulting cost increase.

\begin{proposition}[Per-slot additive mapping gap]\label{prop:fixedK_additive_gap}
Fix \(\FrameLen\ge1\). For every commodity \((c,\phi)\), there exists a finite deterministic constant \(C_{\FrameLen}^{(c,\phi)}\) such that, for every frame \(n\) and every slot \(t\in\mathcal T_n\) with \(A^{(c,\phi)}(t)>0\),
\[
\begin{aligned}
A^{(c,\phi)}(t)
\Psi_t^{(c,\phi)}\!\left(\widehat k_n^{(c,\phi)}\right)
&\le\\
A^{(c,\phi)}(t)
&\min_{k:\,m_k^{(c,\phi)}\in\mathcal M^{(c,\phi)}}
\Psi_t^{(c,\phi)}(k)+
C_{\FrameLen}^{(c,\phi)}.
\end{aligned}
\]
The bound holds uniformly over all slots, queue states, and sample paths. Moreover, \(C_1^{(c,\phi)}=0\) and \(C_{\FrameLen}^{(c,\phi)}=O(\FrameLen-1)\) for each commodity.
\end{proposition}

The complete proof is given in Appendix~\ref{app:virtual}. The next theorem shows that reusing a mapping for any fixed finite frame preserves strong stability of the virtual queues: the expected total backlog averaged over time remains finite for arrival rates in the interior of the capacity region.

\begin{theorem}[Virtual queue stability]\label{theorem:virtual_stable}
Suppose Assumptions~\ref{assumption:wireless_state} and~\ref{assumption:proportional_arrivals} hold and each \(A^{(c,\phi)}(t)\) has finite second moment. Fix \(\FrameLen\ge1\). If \(\FrameLen>1\), additionally let Assumption~\ref{assumption:fixedK_bounded_arrival} hold. Then \(\policyvk\) makes all virtual link and node queues strongly stable for every arrival rate vector in \(\mathrm{int}(\Gamma_{\rm LP})\):
\begin{equation}
\begin{aligned}
\limsup_{T\to\infty}\frac{1}{T}\sum_{t=0}^{T-1}
\Bigg[
\sum_{(u,v)\in\mathcal E}
\mathbb{E}\left[\widetilde Q_{uv}(t)\right]
+\sum_{u\in\mathcal V}
\mathbb{E}\left[\widetilde Q_u(t)\right]
\Bigg]
<\infty.
\end{aligned}
\nonumber
\end{equation}
\end{theorem}

The proof uses the standard quadratic Lyapunov comparison with a stationary randomized policy~\cite{NeelySNO}, as in UMW~\cite{UMW} and UCNC~\cite{UCNC}. Proposition~\ref{prop:fixedK_additive_gap} contributes only a finite additive term to the drift and therefore changes the backlog bound without changing the negative backlog term that establishes strong stability. The complete proof is given in Appendix~\ref{app:virtual}.

Each commodity requires at most one DynVMP call per frame. Increasing \(\FrameLen\) reduces the frequency of mapping computation but may increase the additive mapping gap and virtual queue backlog bound. Under Assumptions~\ref{assumption:wireless_state}--\ref{assumption:fixedK_bounded_arrival}, the same analysis accommodates mapping prices with bounded staleness.

\begin{corollary}[Stale mapping prices]\label{cor:stale}
Suppose that \eqref{eq:fixedK_mapping} uses virtual queue
lengths sampled at most \(d\le\FrameLen\) slots before
the beginning of each frame, where \(d\) is fixed.
The sampled prices are then at most \(\FrameLen+d-1\)
slots old. Under bounded arrivals, virtual queue changes
per slot are bounded, so the argument of
Proposition~\ref{prop:fixedK_additive_gap} yields another
finite additive mapping gap. This changes only the
constant in the Lyapunov drift bound, and the conclusion
of Theorem~\ref{theorem:virtual_stable} remains valid.
\end{corollary}

\subsection{Heuristic Valid Mapping Solver}\label{subsec:heuristic_extension}

DCNC-DAG-H sets \(\FrameLen=1\) and replaces the exact cost minimization in \eqref{eq:fixedK_mapping} with a queue-aware heuristic. The virtual queue updates, max-weight link activation, and NTO scheduling are unchanged.

The heuristic is inspired by RW-MaxMatch~\cite{cheng2011virtual} and first selects execution nodes and then routes the resulting dependencies. It processes the service functions in a fixed topological order. For each function \(A^\phi\), it selects the qualified node that minimizes the sum of its computation price \(r^{A^\phi}\widetilde Q_u(t)\) and the shortest path costs from its already placed predecessors, where dependency \(e\) assigns link \((u,v)\) the price \(\xi^e\widetilde Q_{uv}(t)\). After all functions are placed, each dependency is routed on a minimum-cost path, while a co-located dependency uses the empty path. When all selected dependency endpoints are connected by such
paths, the construction yields a valid mapping.

\section{Dynamic Control of the Physical Network}\label{sec:physical_network}

This section describes the physical queueing system and request execution under DCNC-DAG and establishes physical queue rate stability and throughput optimality.

\subsection{Physical Queueing System}\label{subsec:DCNC_DAG}

The physical system follows the same valid mappings and link activations used by the virtual system, but its workloads are realized progressively as transmissions and computations are completed. Let \(Q_{uv}(t)\) and \(Q_u(t)\) denote the unfinished transmission workload on link \((u,v)\) and the unfinished computation workload at node \(u\), respectively. Each physical workload queue includes the remaining workload currently in service. Transmission on activated links and computation at nodes are work-conserving and nonpreemptive.

\emph{Waiting pools.}
Each computation node maintains a waiting pool to synchronize the inputs required by service functions mapped to that node. The pool consists of waiting components, each associated with a specific request and service function. A waiting component stores the input packets of its request until all inputs required by the corresponding function \(A^\phi\) have arrived. These packets are then consumed to form a processing job of workload \(r^{A^\phi}\), which enters \(Q_u(t)\) at the mapped node \(u\). Let \(W_u^{(c,\phi)}(t)\) denote the total number of packets stored in waiting components of commodity \((c,\phi)\) at node \(u\). A processing job releases its outputs only after completion, and the outputs follow the paths prescribed by the request's mapping.

\emph{Hop grades and NTO.} Let \(\operatorname{hop}(x)\) denote the hop grade of a packet, processing job, or waiting component \(x\). Each source packet has grade zero, and completing a link transmission increases the packet grade by one. A function execution also contributes one hop. If a processing job \(j\) is formed from the input packet set \(\mathcal B_j\), then \(\operatorname{hop}(j)=\max_{b\in\mathcal B_j}\operatorname{hop}(b)\), and each output packet \(b'\) of \(j\) has grade \(\operatorname{hop}(b')=\operatorname{hop}(j)+1\). Transmissions along empty paths do not change the grade. An incomplete waiting component takes the grade of the processing job it will form. Given the request's valid mapping, these recursive rules uniquely determine the grade of each object throughout its execution, and grades remain unchanged while objects wait or receive service. Because service DAGs are finite and mapped paths are simple, all hop grades are bounded. Whenever an activated link or computation node is ready to start a new service, NTO scheduling~\cite{NTO_origin} selects an eligible packet or processing job with the smallest grade. Service is nonpreemptive, so the selected packet or job remains in service until completion; consequently, at most one partially served packet or job remains at each link or node across a slot boundary.

At the beginning of slot \(t\), newly arrived requests and any packets or jobs released at the end of slot \(t-1\) become eligible for service. Transfers over empty paths are resolved before service begins. Activated links and computation nodes serve eligible packets and processing jobs under NTO, subject to their capacities in slot \(t\). Successors generated by service completions during slot \(t\) are released at the end of the slot and become eligible in slot \(t+1\).

Algorithm~\ref{alg:DCNC-DAG} presents the DCNC-DAG algorithm. Mapping computation and reuse follow Section~\ref{subsec:fixedK_policy}. Setting \(\FrameLen=1\) and replacing the exact mapping solver with the heuristic of Section~\ref{subsec:heuristic_extension} yields DCNC-DAG-H.

\begin{algorithm}
    \caption{DCNC-DAG}
    \begin{algorithmic}[1]
        \STATE Initialize all virtual queues, physical queues, and waiting pools to zero.
        \FOR{each slot \(t\)}
        \STATE Let \(n=\lfloor t/\FrameLen\rfloor\).
        \STATE \textbf{Valid mapping.} For each commodity \((c,\phi)\) with \(A^{(c,\phi)}(t)>0\), compute \(\widehat k_n^{(c,\phi)}\) from \eqref{eq:fixedK_mapping} if no mapping is stored for frame \(n\); otherwise reuse the stored mapping. Assign the arriving requests accordingly, form the virtual queue arrivals, and generate the source packets.
        \STATE \textbf{Link activation.} Select \(\mathbf s^*(t)\) according to \eqref{eq:max_weight_activation}.
        \STATE \textbf{Physical scheduling.} Serve eligible packets on link \((u,v)\) with capacity \(\mu_{uv}(t)s_{uv}^{*}(t)\) and eligible processing jobs at node \(u\) with capacity \(\mu_u\), using nonpreemptive NTO. Successors released by completed service become eligible in slot \(t+1\).
        \STATE \textbf{Virtual queue update.} Update the virtual queues according to \eqref{eq:virtual_Quv_evolution}--\eqref{eq:virtual_Qu_evolution} using the selected mappings and \(\mathbf{s}^{*}(t)\).
        \ENDFOR
    \end{algorithmic}
    \label{alg:DCNC-DAG}
\end{algorithm}

\subsection{Physical Network Stability}\label{subsec:NTO_stability}

Virtual workloads are charged at request arrival but realized progressively in the physical system, so virtual queue stability does not directly imply physical queue rate stability. We build on the NTO hop induction in~\cite{NTO_origin} and the virtual to physical arguments in UMW~\cite{UMW} and UCNC~\cite{UCNC}, which control progressively released transmission and computation work by induction over hop grades.

For service DAGs, the additional difficulty is that packets can remain in waiting components without contributing to a link or computation queue, and a missing input may itself depend on an incomplete upstream merge. Let \(Q_r^h(t)\) be the physical workload of grade \(h\) at resource \(r\), let \(Q^h(t)=\sum_r Q_r^h(t)\), and let \(W^h(t)\) count packets in waiting components of grade \(h\). Superscripts \(\le h\) and \(<h\) denote sums over the corresponding grades, and \(H_{\max}\) bounds all grades. The following lemma incorporates this synchronization backlog into the induction.

\begin{lemma}[Waiting pool bound]\label{lemma:witness_charging_main}
For every \(h<H_{\max}\), one missing input of each nonempty waiting component of grade \(h+1\) can be traced backward within the same request to either an unfinished packet or processing job of grade at most \(h\), or an upstream waiting component of grade at most \(h\). The traces can be chosen so that the number terminating at any one such object is bounded by a constant independent of \(t\). Consequently, for some finite \(\eta_{\rm w}\) independent of \(t\),
\[
    W^{h+1}(t)
    \le
    \eta_{\rm w}\bigl(1+Q^{\le h}(t)+W^{\le h}(t)\bigr).
\]
\end{lemma}

The proof traces one missing input of each waiting component backward within the same request. If the packet has already been generated, an unfinished transmission on its prescribed path provides a lower grade object. Otherwise, the packet is an intermediate output, and the trace moves to the predecessor function that produces it. If the corresponding processing job has formed and remains unfinished, the trace stops at that job. Otherwise, it stops at the function's nonempty waiting component or continues through one of its missing inputs. Each continuation moves to a predecessor function, so acyclicity and the availability of all source packets at admission guarantee termination. Because the trace never leaves the request and each request contains only a bounded number of waiting components and packets, only a bounded number of such traces can terminate at the same unfinished object. This recursive tracing bounds the waiting backlog at merge points, allowing the physical stability analysis to account for merge synchronization. Appendix~\ref{app:physical} gives the complete argument.

The next theorem establishes that, under NTO scheduling, both unfinished workloads and packets waiting for synchronization grow sublinearly with time, ensuring rate stability of the physical queues and waiting pools.

\begin{theorem}[Physical network rate stability]\label{theorem:physical_stable}
Fix \(\FrameLen\ge1\) and suppose Assumptions~\ref{assumption:wireless_state}--\ref{assumption:fixedK_bounded_arrival} hold. Under DCNC-DAG, for every \(\boldsymbol\Lambda\in\mathrm{int}(\Gamma_{\rm LP})\), with probability one,
\[
\begin{aligned}
    \lim_{t\to\infty}\frac{Q_{uv}(t)}{t}&=0 && \forall (u,v)\in\mathcal E,\\
    \lim_{t\to\infty}\frac{Q_u(t)}{t}&=0 && \forall u\in\mathcal V,\\
    \lim_{t\to\infty}\frac{W_u^{(c,\phi)}(t)}{t}&=0 && \forall u\in\mathcal V,\ (c,\phi)\in\mathcal K.
\end{aligned}
\]
\end{theorem}

Appendix~\ref{app:physical} first obtains almost sure rate stability of the virtual queues and then applies the standard interval workload and NTO busy period arguments to bound physical work of each grade. The physical queue arrival bound accounts for unfinished work and waiting packets present at the beginning of an interval, while Lemma~\ref{lemma:witness_charging_main} bounds the waiting backlog at the next hop grade. Alternating these two bounds over the finitely many hop grades establishes Theorem~\ref{theorem:physical_stable}.

\begin{corollary}[Throughput optimality]\label{cor:throughput}
Under the conditions of Theorem~\ref{theorem:physical_stable}, let \(I(t)\) be the number of admitted but incomplete requests. Then, with probability one, \(\lim_{t\to\infty}\frac{I(t)}{t}=0\) and
\[
    \lim_{t\to\infty}\frac{\kappa^{(c,\phi)}(t)}{t}=\lambda^{(c,\phi)}
\]
for every commodity. Hence, for every fixed \(\FrameLen\), DCNC-DAG supports every arrival rate vector in \(\mathrm{int}(\Gamma_{\rm LP})\) and is throughput-optimal.
\end{corollary}

The complete proof is given in Appendix~\ref{app:throughput_proof}.

Although \(\Gamma_{\rm LP}\) contains no explicit synchronization constraints, Corollary~\ref{cor:throughput} shows that every arrival rate vector in its interior is achievable while the waiting backlog at every node remains rate stable. Thus synchronization does not shrink the achievable capacity region.

The proof relies on NTO's priority for smaller hop grades. Section~\ref{subsec:simulation_FIFO} illustrates its effect on a re-entrant instance, where NTO keeps the backlog of incomplete requests small while FIFO exhibits sustained backlog growth under identical arrivals, implying instability.

\section{Simulation Results}\label{sec:simulation}

We compare completed throughput with the LP capacity boundary, evaluate queue-aware mapping across wireless topologies and larger service DAGs, and examine the effects of frame length and scheduling. The simulation code and experimental data are publicly available.\footnote{\url{https://github.com/Peng0802490/dcnc-dag-reproducibility/releases/tag/v0.1.1}}

\subsection{Simulation Setup}\label{subsec:simulation_setup}
\emph{Networks, service DAGs, and traffic.} Table~\ref{tab:sim_topologies} lists the four wireless network topologies. Node and link capacities are drawn independently and uniformly from \(\{2,\ldots,10\}\) and then fixed. Each undirected link gives two directed arcs. Dynamic arcs are independently ON with probability \(0.8\) in each slot and direction; all others remain ON. The wireless experiments use primary interference. Compared policies share the same arrival and wireless sample paths. The LP capacity boundary experiment uses bounded arrivals, whereas the mapping policy and frame length experiments use Poisson arrivals. Appendix~\ref{app:simulation_config} provides the remaining simulation configurations.

\begin{table}[htbp]
\centering
\caption{Wireless network topologies used in the experiments.}
\label{tab:sim_topologies}
\footnotesize
\setlength{\tabcolsep}{4pt}
\begin{tabular}{@{}lrrrr@{}}
\hline
Topology & \(|\mathcal V|\) & Links & \shortstack{Qualified\\nodes} & \shortstack{Dynamic\\links} \\
\hline
Milan~\cite{MilanTopology}        & 30 & 35  & 11 & 4  \\
Random                            & 50 & 50  & 16 & 6  \\
RoofNet~\cite{Bicket2005RoofNet}  & 38 & 131 & 10 & 35 \\
FlockLab~\cite{Jacob2019FlockLab} & 26 & 171 & 10 & 66 \\
\hline
\end{tabular}
\tabnote{Link counts are undirected. The numbers of qualified nodes refer to the three random service DAG experiments; candidate sets for other experiments are specified in Appendix~\ref{app:simulation_config}.}
\end{table}

\emph{Compared policies.} We compare DCNC-DAG with \(\FrameLen=1\) unless stated otherwise, DCNC-DAG-H, Closest-to-Source (CTS), and Randomized. CTS processes functions in topological order and places each function at a qualified node minimizing the sum of hop distances from its already placed predecessors; each dependency then follows a shortest path. Randomized first selects a qualified computation node for each service function uniformly at random, and then routes each dependency along a shortest path between the selected nodes. These policies use the same admission, link activation, and NTO scheduling and differ only in the valid mapping computation rule. The MoE experiments on larger service DAGs use two additional placement baselines defined in Section~\ref{subsubsec:large_service_dags}.

\emph{Performance metrics.} We measure request completion delay from arrival until all sink outputs are delivered and average it over completed requests. The empirical supported rate \(\hat\lambda\) is the largest tested rate for which the mean completion ratio across simulation seeds is at least \(0.95\). For each simulation seed, the completion ratio is the total number of completed requests divided by the total number of arrived requests across all commodities over the simulation horizon. This criterion may accept slightly overloaded rates. Conversely, requests that remain incomplete at time \(T\) reduce the completion ratio and may lower \(\hat{\lambda}\). Thus, \(\hat{\lambda}\) is an empirical measure subject to rate-grid and finite-horizon effects, and is a proxy for the stability boundary.

\subsection{Comparison With the LP Capacity Boundary}\label{subsec:throughput_boundary_validation}

We first compare completed throughput with the LP capacity boundary on a wireless fork-join instance for which all valid mappings and wireless states can be enumerated.

\emph{Instance and LP boundary.} The service DAG is \(A^{\phi}\!\to\!B^{\phi},A^{\phi}\!\to\!C^{\phi},B^{\phi}\!\to\!D^{\phi},C^{\phi}\!\to\!D^{\phi}\), so the job at \(D^{\phi}\) must synchronize the two branches. Function \(C^{\phi}\) has unit computation workload and may execute at either of two nodes with capacity \(0.5\), while the remaining functions have fixed execution nodes and do not constrain the capacity boundary. Hence there are exactly two valid mappings, \(m_1\) and \(m_2\), according to the placement of \(C^{\phi}\). Either mapping alone supports at most \(0.5\) requests per slot. Enumerating both mappings, all \(2^4\) wireless states, and all feasible activations gives \(\lambda^{\star}=1\), attained only by \(f_{m_1}=f_{m_2}=0.5\). Appendix~\ref{app:simulation_config} gives the complete instance configuration.

\emph{Results.}
Table~\ref{tab:throughput_boundary_validation} reports DCNC-DAG over \(10^5\) slots. Below the LP boundary, completed throughput closely follows the offered rate. Above the boundary, completed throughput saturates at \(1\) and the completion ratio decreases. Restricting the policy to a single mapping limits completed throughput to \(0.5\) requests per slot. These results show that completed throughput remains consistent with the LP capacity boundary despite synchronization at the merge and that achieving the boundary requires a convex combination of valid mappings, implemented by assigning half of the requests to each mapping in the long run.

\begin{table}[htbp]
\centering
\caption{Completed throughput and realized mapping share on the fork-join instance with LP capacity boundary \(\lambda^{\star}=1\). The offered arrival rate is \(\rho\) requests/slot.}
\label{tab:throughput_boundary_validation}
\footnotesize
\begin{tabular}{rrrr}
\hline
\(\rho\) & \(\lambda_{\rm out}\) & Comp.\ ratio & Share \(m_1\) \\
\hline
0.95 & 0.950 & 1.000 & 0.507 \\
0.98 & 0.980 & 1.000 & 0.504 \\
0.99 & 0.990 & 1.000 & 0.502 \\
1.01 & 1.000 & 0.990 & 0.500 \\
1.03 & 1.000 & 0.970 & 0.500 \\
1.05 & 1.000 & 0.952 & 0.500 \\
1.10 & 1.000 & 0.909 & 0.500 \\
\hline
\end{tabular}
\tabnote{\(\lambda_{\rm out}\) denotes completed throughput, and Share \(m_1\) denotes the fraction of arrivals assigned to \(m_1\). Values are means over five seeds; all standard errors are below \(5\times10^{-4}\).}
\end{table}

\subsection{Performance of DCNC-DAG and DCNC-DAG-H}\label{subsec:mapping_policy_evaluation}\label{subsec:network_level_eval}

\subsubsection{Comparison With Queue-Blind Mapping Policies}\label{subsubsec:queue_blind_mapping}\label{subsubsec:sim_ad_task}\label{subsubsec:sim_random_dag_topology}

\begin{figure}[htbp]
    \centering
    \subfloat[Milan topology.]{%
        \includegraphics[width=0.48\linewidth]{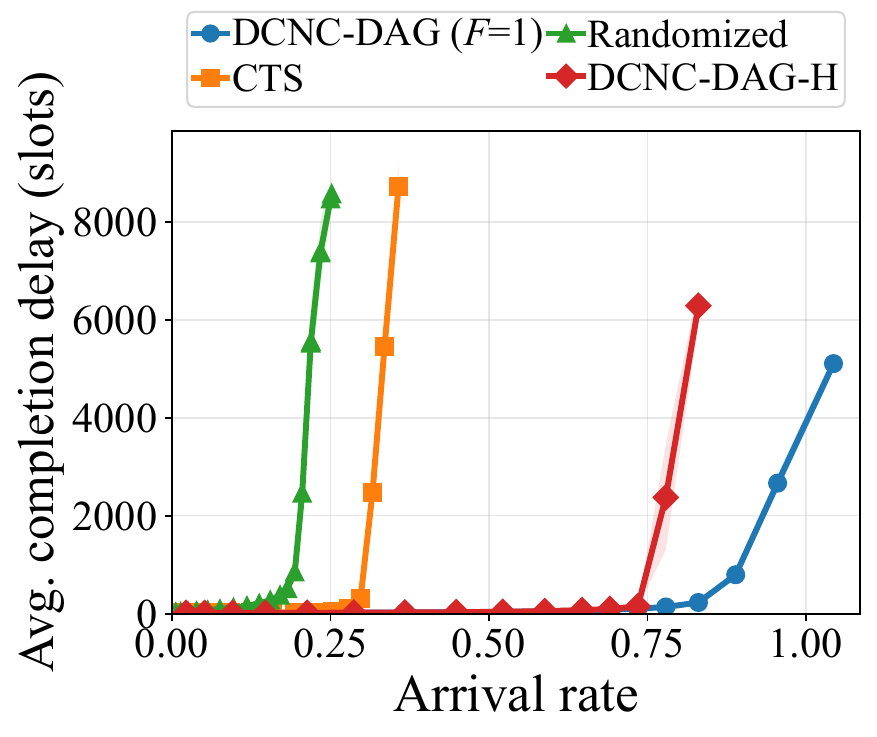}%
        \label{subfig:sim_ad_milan}}
    \hfill
    \subfloat[Random topology.]{%
        \includegraphics[width=0.48\linewidth]{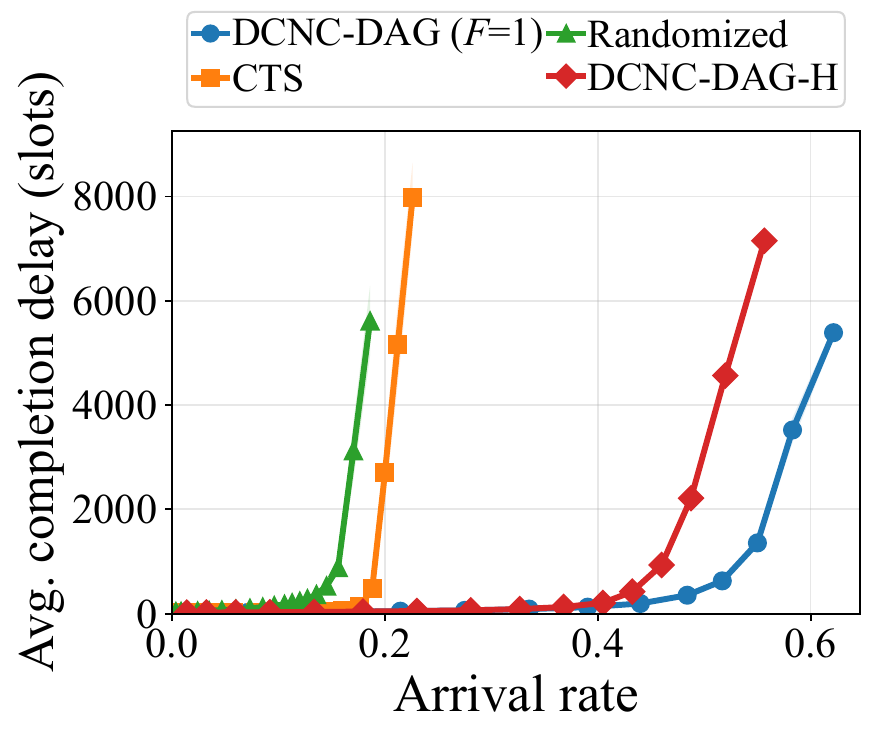}%
        \label{subfig:sim_ad_random}}
    \caption{Completion delay for the autonomous driving service DAG.}
    \label{fig:sim_ad_task}
\end{figure}

\emph{Autonomous driving service DAG.} Fig.~\ref{fig:sim_ad_task} evaluates the autonomous driving service DAG of Fig.~\ref{fig:service_DAG} on the Milan and Random topologies under the wireless mapping configuration with a simulation horizon of \(2\times10^4\) slots. DCNC-DAG achieves empirical supported rates of \(0.83\) and \(0.52\), respectively, compared with \(0.30\) and \(0.19\) for the better queue-blind baseline. DCNC-DAG-H reaches \(0.73\) and \(0.43\), which are approximately \(88\%\) and \(84\%\) of the corresponding values achieved by DCNC-DAG. The delay differences therefore reflect mapping decisions: DCNC-DAG and DCNC-DAG-H use virtual queue length information to avoid congested resources, whereas CTS and Randomized do not.

\begin{figure}[htbp]
    \centering
    \subfloat[Milan topology.]{%
        \includegraphics[width=0.48\linewidth]{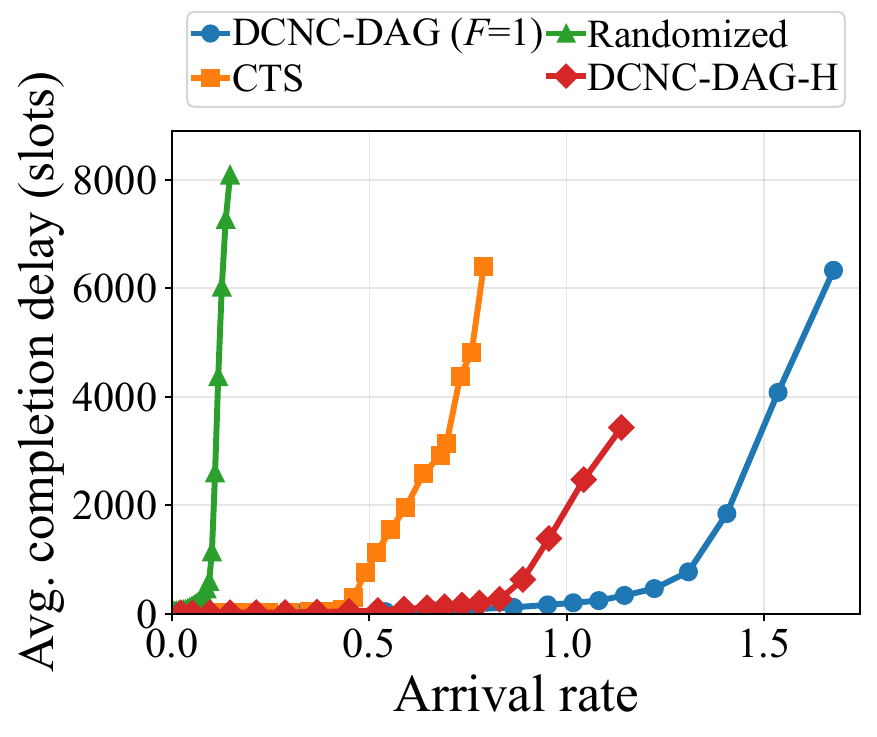}%
        \label{subfig:sim_randomdag_milan}}
    \hfill
    \subfloat[Random topology.]{%
        \includegraphics[width=0.48\linewidth]{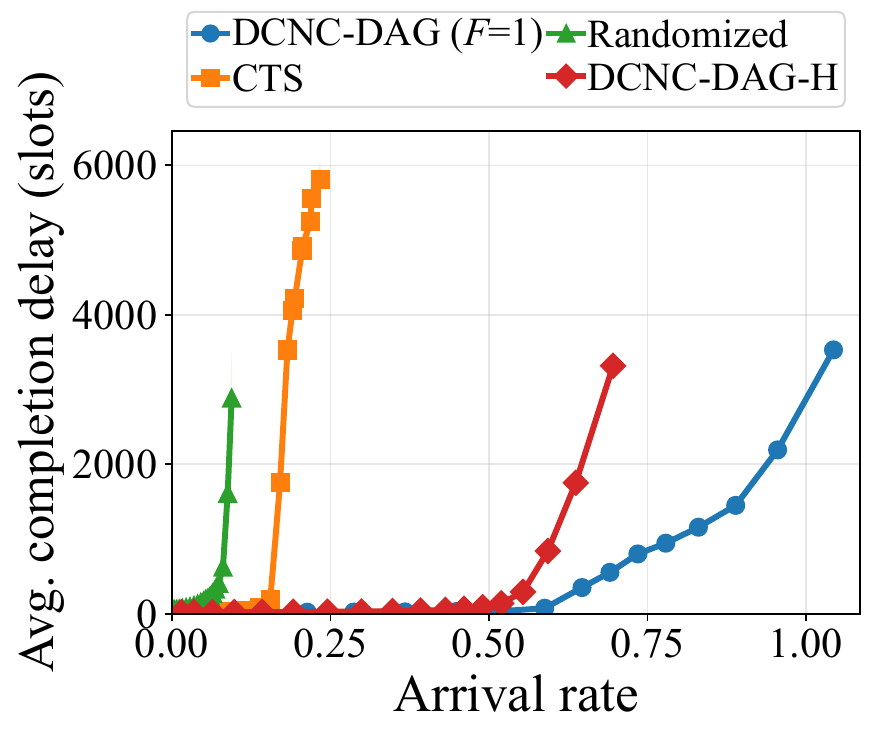}%
        \label{subfig:sim_randomdag_random}}
    \par\smallskip
    \subfloat[RoofNet topology.]{%
        \includegraphics[width=0.48\linewidth]{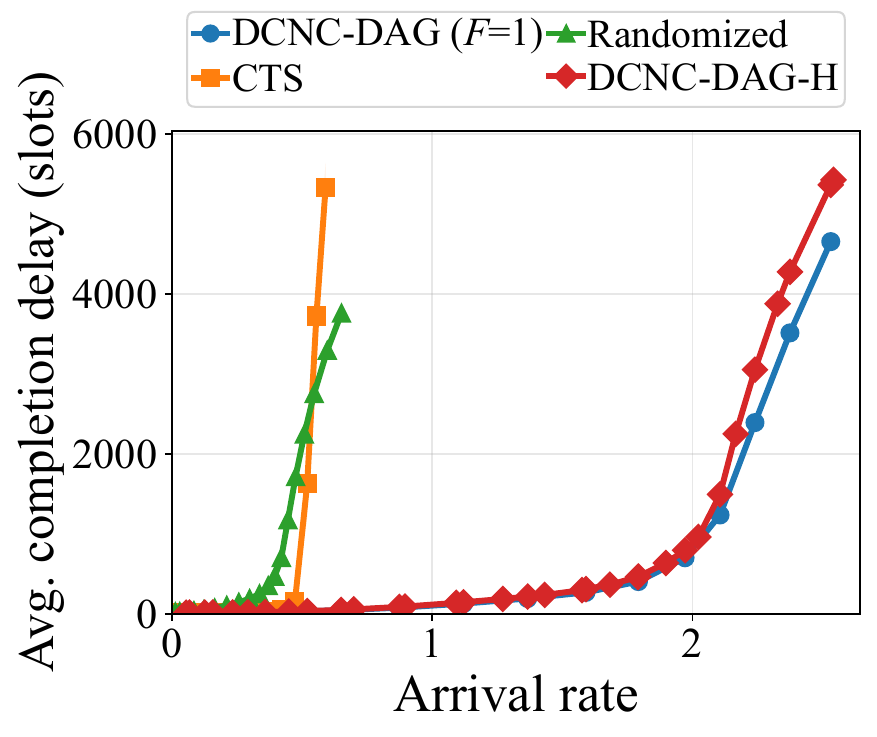}%
        \label{subfig:sim_randomdag_roofnet}}
    \hfill
    \subfloat[FlockLab topology.]{%
        \includegraphics[width=0.48\linewidth]{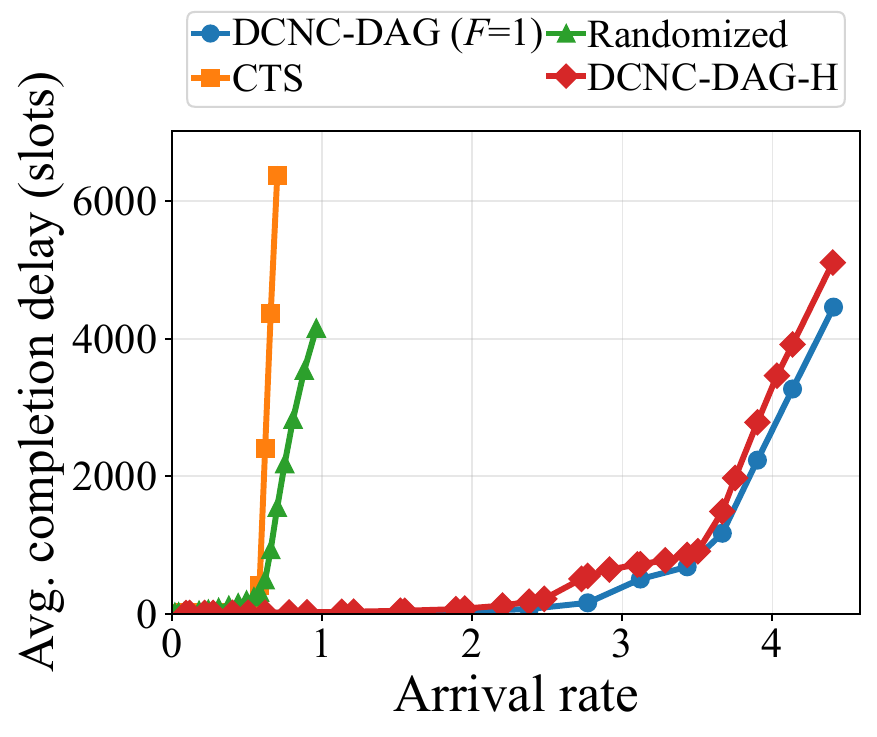}%
        \label{subfig:sim_randomdag_flocklab}}
    \caption{Completion delay for three random service DAGs across the four wireless topologies.}
    \label{fig:sim_random_dag_topology}
\end{figure}

\emph{Random service DAGs and wireless topologies.} Table~\ref{tab:sim_randomdag_rates} and Fig.~\ref{fig:sim_random_dag_topology} evaluate three six-function service DAGs across all four topologies. DCNC-DAG achieves empirical supported rates \(2.7\times\)--\(5.5\times\) those of the better queue-blind baseline. DCNC-DAG-H achieves approximately \(68\%\) and \(86\%\) of the corresponding DCNC-DAG values on Milan and the Random topology, while the two queue-aware policies are closer on the denser topologies. These results show the benefit of queue-aware mapping and the additional gain from exact cost minimization, particularly when path diversity is limited.

\begin{table}[htbp]
\centering
\caption{Empirical supported rates \(\hat\lambda\) (requests/slot) for three random service DAGs across the four wireless topologies.}
\label{tab:sim_randomdag_rates}
\footnotesize
\setlength{\tabcolsep}{6pt}
\begin{tabular}{@{}lrr@{}}
\hline
Topology & DCNC-DAG & \shortstack{Better queue-blind policy} \\
\hline
Milan    & 1.22 & 0.46 \\
Random   & 0.65 & 0.15 \\
RoofNet  & 1.97 & 0.47 \\
FlockLab & 3.43 & 0.62 \\
\hline
\end{tabular}
\end{table}

\subsubsection{Performance of DCNC-DAG-H on Large Service DAGs}\label{subsubsec:large_service_dags}\label{subsec:simu_AI}

The following experiments evaluate DCNC-DAG-H on MoE and workflow service DAGs with 20--31 functions. The MoE experiments use the Milan topology, while the workflow experiments use the Random topology. Table~\ref{tab:sim_largedag_rates} summarizes the corresponding empirical supported rates.

\begin{table}[htbp]
\centering
\caption{Empirical supported rates \(\hat\lambda\) for the large service DAGs.}
\label{tab:sim_largedag_rates}
\footnotesize
\setlength{\tabcolsep}{2pt}
\begin{tabular}{@{}lrrrl@{}}
\hline
Setting & DCNC-DAG-H & Lazarus & FlexMoE & Gain \\
\hline
\multicolumn{5}{l}{\emph{Skewed MoE service DAGs}} \\
MoE Mixtral   & 1.16 & 0.38 & 0.44 & \(2.64\times\) \\
MoE Llama     & 1.02 & 0.30 & 0.38 & \(2.68\times\) \\
\hline
Setting & DCNC-DAG-H & CTS & Randomized & Gain \\
\hline
\multicolumn{5}{l}{\emph{Workflow service DAGs}} \\
SRA-Search-22 & 0.14 & 0.11  & 0.03 & \(1.27\times\) \\
SIPHT-31      & 0.30 & 0.30  & 0.07 & \(1.00\times\) \\
\hline
\end{tabular}
\tabnote{Gain is relative to the better baseline in each row. Lazarus and FlexMoE denote the corresponding inspired placement rules used in the MoE experiments.}
\end{table}

\begin{figure}[htbp]
    \centering
    \subfloat[Mixtral-skewed.]{%
        \includegraphics[width=0.48\linewidth]{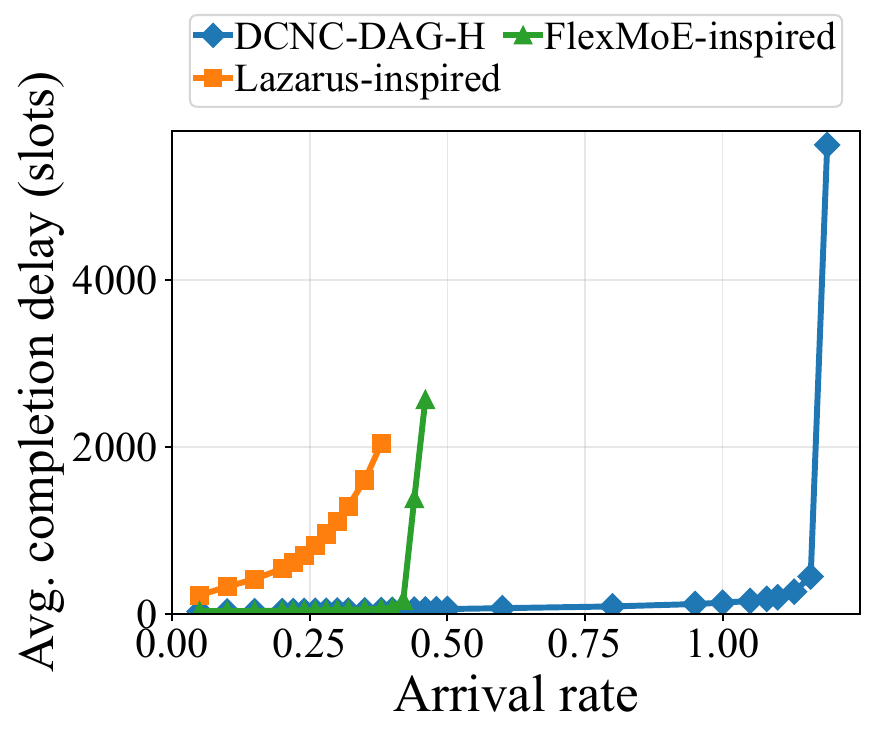}%
        \label{subfig:moe_mixtral_skewed}}
    \hfill
    \subfloat[Llama-skewed.]{%
        \includegraphics[width=0.48\linewidth]{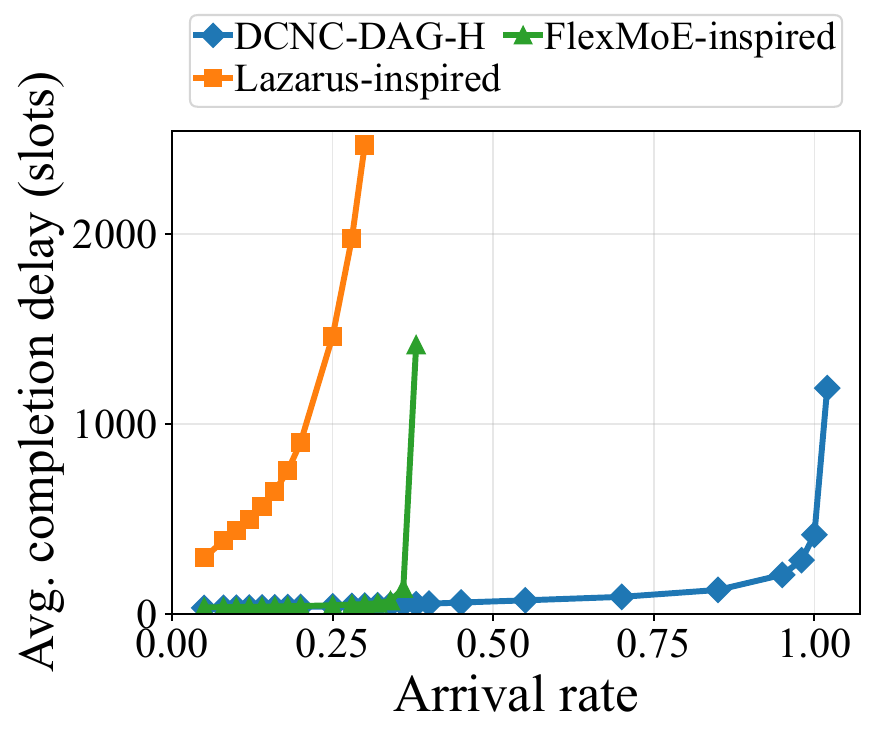}%
        \label{subfig:moe_llama_skewed}}
    \par\smallskip
    \subfloat[SRA-Search-22.]{%
        \includegraphics[width=0.48\linewidth]{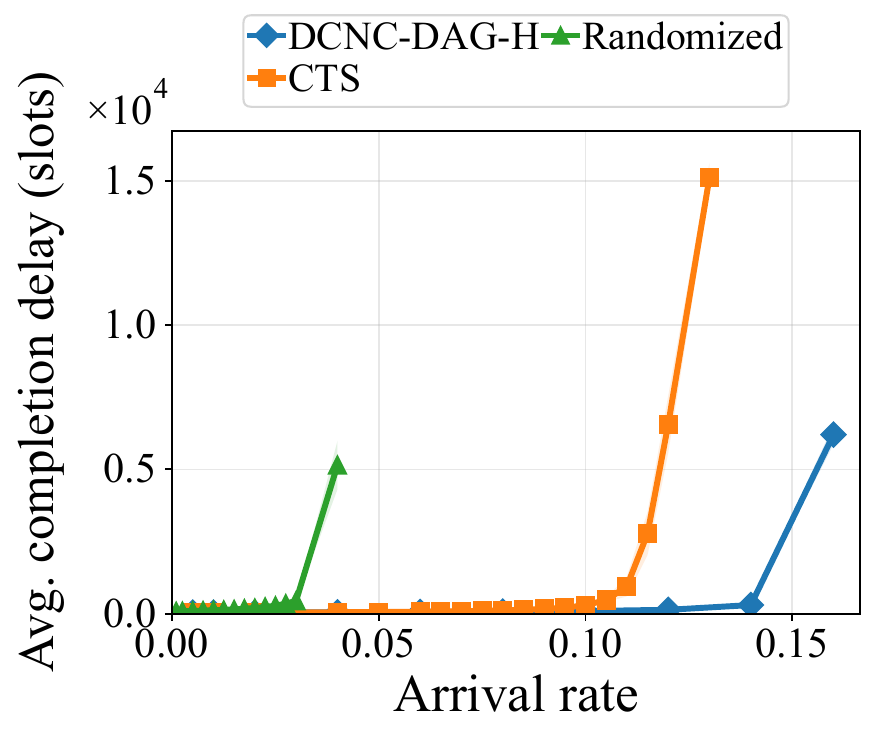}%
        \label{subfig:sim_workflow_sra}}
    \hfill
    \subfloat[SIPHT-31.]{%
        \includegraphics[width=0.48\linewidth]{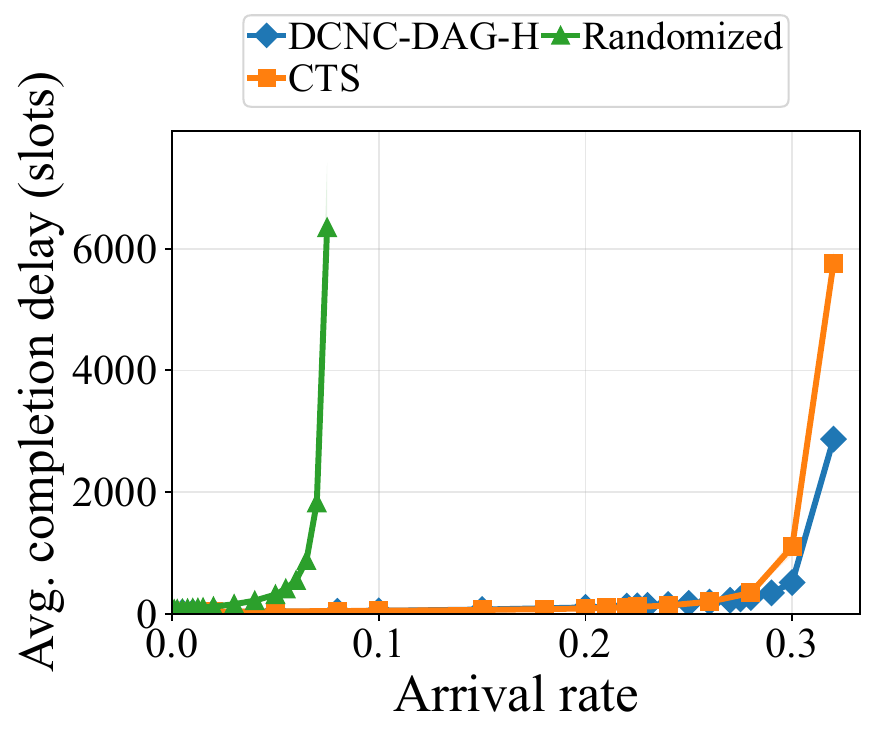}%
        \label{subfig:sim_workflow_sipht}}
    \caption{Completion delay for the skewed MoE and workflow-derived service DAGs.}
    \label{fig:sim_large_service_dags}
\end{figure}

\emph{MoE service DAGs.} Each expert family is modeled as one service function with several qualified GPU locations, and each request executes the function once. The workloads are derived from representative Mixtral 8\(\times\)7B~\cite{jiang2024mixtral} and Llama-4-Maverick~\cite{llama4maverick2025} operator magnitudes with skewed expert demand. The qualified computation nodes in the Milan topology are treated as edge servers equipped with GPUs. The Lazarus-inspired baseline~\cite{wu2024lazarus} selects the qualified GPU with the smallest physical backlog, while the FlexMoE-inspired baseline~\cite{nie2023flexmoe} uses a fixed rotating order; after placement, both use the same minimum-cost dependency routing rule and NTO scheduling as DCNC-DAG-H. Table~\ref{tab:sim_largedag_rates} shows that DCNC-DAG-H achieves empirical supported rates \(2.64\times\) and \(2.68\times\) those of the better MoE-inspired baseline for Mixtral and Llama, respectively. Fig.~\ref{fig:sim_large_service_dags}(a)--(b) shows the corresponding delay curves. Appendix~\ref{app:simulation_config} gives the workload construction, candidate sets, and simulation configuration.

\emph{Scientific data analytics workflows.} SRA-Search-22 and SIPHT-31 are bioinformatics workflows based on published WfCommons/Pegasus instances~\cite{WfCommons2022}. These workflows provide larger DAGs with branching and merging dependencies for evaluating DCNC-DAG-H. Table~\ref{tab:sim_largedag_rates} shows that DCNC-DAG-H achieves an empirical supported rate \(1.27\times\) that of the better queue-blind baseline on SRA-Search-22. On SIPHT-31, DCNC-DAG-H and CTS both support \(0.30\) requests per slot. Fig.~\ref{fig:sim_large_service_dags}(c)--(d) compares their request completion delays. At \(\lambda=0.30\) on SIPHT-31, the mean completed-request delay is \(510.30\) slots for DCNC-DAG-H and \(1098.67\) slots for CTS, a reduction of approximately \(53.6\%\). Appendix~\ref{app:simulation_config} gives the network, workload conversion, and simulation configuration.

\subsection{Frame Length and Scheduling}\label{subsec:frame_and_scheduling}

\subsubsection{Effect of the Frame Length \(\FrameLen\)}\label{subsubsec:sim_frame_length}\label{subsec:sim_frame_length}

\begin{figure}[htbp]
    \centering
    \subfloat[Milan topology.]{%
        \includegraphics[width=0.48\linewidth]{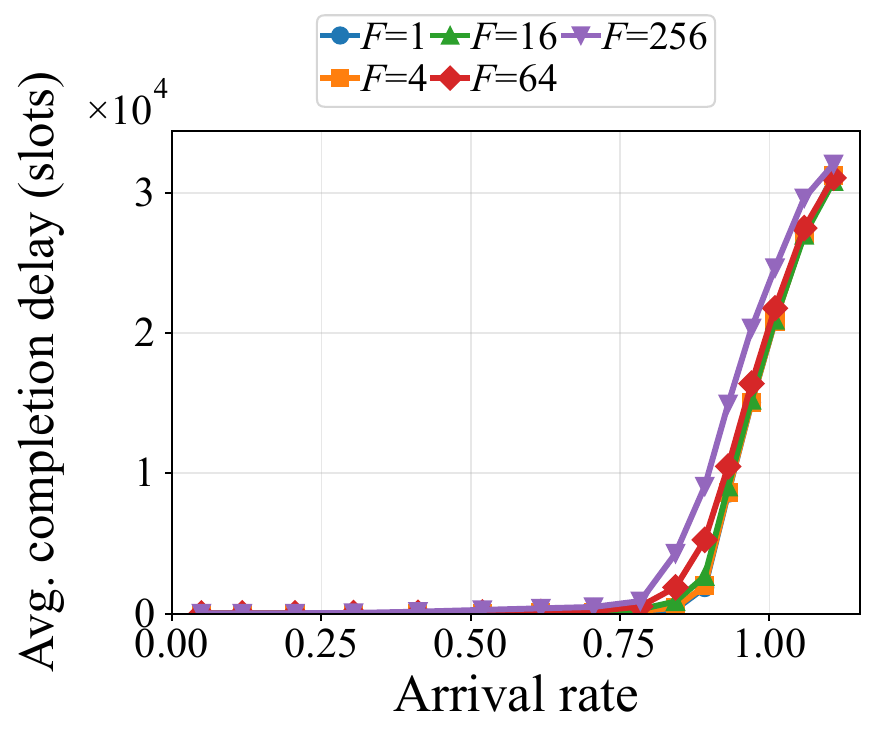}%
        \label{subfig:sim_frame_milan}}
    \hfill
    \subfloat[Random topology.]{%
        \includegraphics[width=0.48\linewidth]{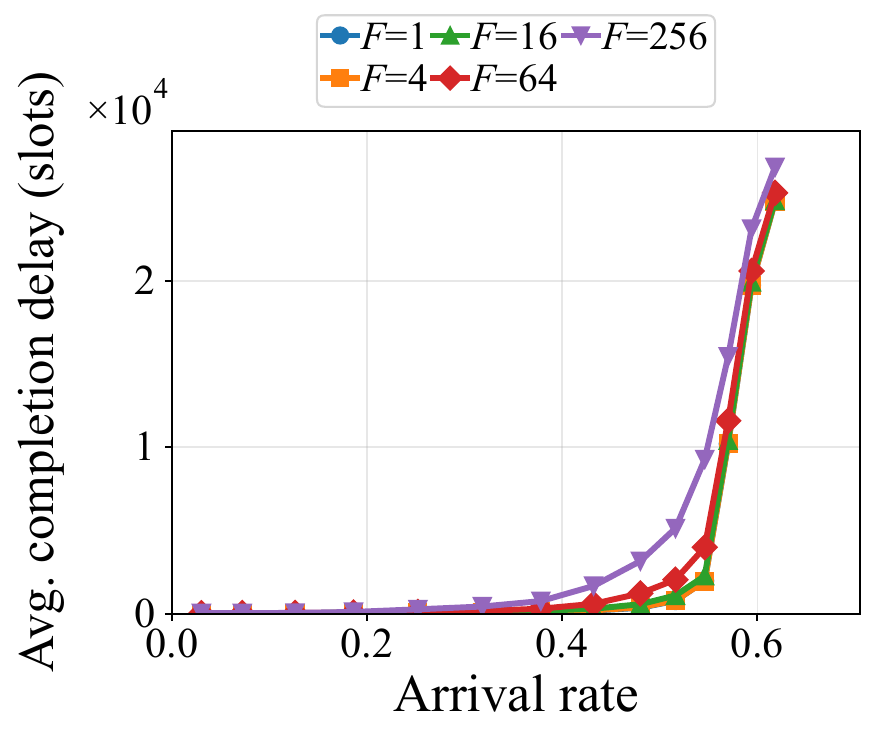}%
        \label{subfig:sim_frame_random}}
    \caption{Completion delay for \(\FrameLen\in\{1,4,16,64,256\}\) on the autonomous driving service DAG. Curves show means over three seeds; shaded bands indicate one standard error and are visually subtle due to their narrow widths.}
    \label{fig:sim_frame_length}
\end{figure}

We vary \(\FrameLen\in\{1,4,16,64,256\}\) for DCNC-DAG with exact mappings on the autonomous driving service DAG, using the Milan and Random topologies considered in Section~\ref{subsubsec:queue_blind_mapping}. Appendix~\ref{app:simulation_config} gives the simulation configuration and common rate grids. Table~\ref{tab:sim_frame_overhead} reports the empirical supported rate \(\hat\lambda\) and exact mapping overhead, and Fig.~\ref{fig:sim_frame_length} reports the corresponding delay curves.

\begin{table}[htbp]
\centering
\caption{Effect of frame length on the empirical supported rate and exact mapping overhead.}
\label{tab:sim_frame_overhead}
\footnotesize
\setlength{\tabcolsep}{2.1pt}
\renewcommand{\arraystretch}{1.05}
\begin{tabular*}{\columnwidth}{@{\extracolsep{\fill}}crrrrrr@{}}
\toprule
& \multicolumn{3}{c}{Milan} & \multicolumn{3}{c}{Random} \\
\cmidrule(lr){2-4}\cmidrule(l){5-7}
\(\FrameLen\) & \(\hat\lambda\) & \shortstack{Map.\\calls} & \shortstack{Map. time/\\slot (s)}
          & \(\hat\lambda\) & \shortstack{Map.\\calls} & \shortstack{Map. time/\\slot (s)} \\
\midrule
1   & 0.89 & 36302.9 & 2.825  & 0.55 & 25432.3 & 11.028 \\
4   & 0.89 & 18365.0 & 1.626  & 0.55 & 16273.9 &  7.254 \\
16  & 0.89 &  5944.1 & 0.528  & 0.55 &  5931.1 &  2.991 \\
64  & 0.84 &  1562.5 & 0.139  & 0.55 &  1562.5 &  0.852 \\
256 & 0.78 &   391.0 & 0.0349 & 0.48 &   391.0 &  0.209 \\
\bottomrule
\end{tabular*}
\tabnote{\(\hat\lambda\) uses the threshold/delay rate grids. Map. calls is the number of exact mapping invocations per commodity over \(10^5\) slots; Map. time/slot is their cumulative wall-clock time divided by that horizon. Both overhead measures are averaged over three seeds on separate five-point rate grids (Appendix~\ref{app:sim_configurations}).}
\end{table}

As \(\FrameLen\) increases from \(1\) to \(256\), the number of exact mapping calls per commodity decreases by \(98.9\%\) on Milan and \(98.5\%\) on the Random topology, while the amortized mapping computation time per slot decreases by \(98.8\%\) and \(98.1\%\), respectively. Since any fixed finite \(\FrameLen\) preserves throughput optimality, increasing the frame length does not change the supported throughput region. The measured \(\hat{\lambda}\) differs slightly across frame lengths, primarily due to finite-horizon effects. Completion delay increases with \(\FrameLen\), particularly at high offered rates. Thus, frame reuse reduces mapping computation at the cost of increased completion delay, without changing the supported throughput region.

\subsubsection{Comparison of FIFO and NTO Scheduling}\label{subsubsec:simulation_FIFO}\label{subsec:simulation_FIFO}

\emph{Re-entrant instance and configuration.} To examine the role of NTO, we embed the classical re-entrant line with two stations~\cite{Bramson_reentrant_line} as a six-function service chain on two computation nodes, each with unit capacity. The per-request computation loads are \(1.00\) at \(u_1\) and \(1.02\) at \(v_1\), so \(\lambda^{\star}=1/1.02\approx0.9804\). We use Bernoulli arrivals at \(\lambda=0.80\), start empty, and compare FIFO with NTO under identical arrivals, mappings, capacities, and nonpreemptive work-conserving service; only the service order differs.

\begin{figure}[htbp]
    \centering
    \includegraphics[width=\linewidth]{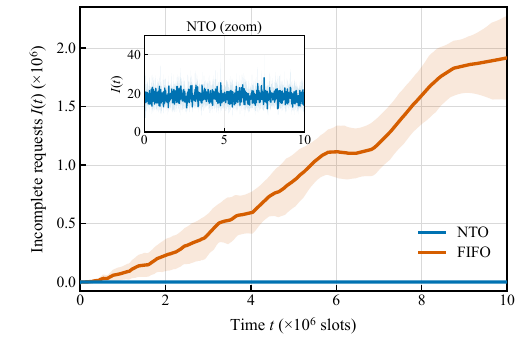}
    \caption{Backlog of incomplete requests under NTO and FIFO on the re-entrant line at \(\lambda=0.80<\lambda^{\star}\). Curves are means over ten seeds; shading gives 95\% confidence intervals of the mean; the inset enlarges NTO.}
    \label{fig:FIFO_simulation}
\end{figure}

\emph{Results and scope.} Fig.~\ref{fig:FIFO_simulation} shows the backlog \(I(t)\) of incomplete requests. NTO keeps it on the order of tens of requests, whereas FIFO accumulates about \(1.9\times10^6\) by \(10^7\) slots despite \(\lambda<\lambda^{\star}\); the inset enlarges the NTO trajectory. Thus, scheduling can lead to large differences in incomplete request backlog over the simulated horizon
in this re-entrant system.

\FloatBarrier

\section{Conclusion}\label{sec:conclusion}
We characterize the capacity region of wireless distributed computing networks serving service DAGs through the valid mapping abstraction and show, through the waiting pool analysis, that merge synchronization does not shrink this capacity region. We develop DCNC-DAG and prove that, with exact mappings and bounded arrivals, it is throughput-optimal for every fixed frame length, including when mapping prices are boundedly stale. We also develop DCNC-DAG-H, which applies queue-aware mapping to larger workflow and AI service DAGs. Our experiments show consistency with the LP boundary, demonstrate gains over queue-blind and domain-specific baselines, quantify the tradeoff between mapping computation and delay under frame reuse, and illustrate the effect of scheduling on re-entrant execution.

\appendices
\setcounter{equation}{7}
\setcounter{table}{6}
\setcounter{figure}{6}
\setcounter{lemma}{2}

\section{Capacity Region Characterization}\label{app:capacity}

\subsection{Valid Mapping Decomposition}

\begin{proof}[Proof of Lemma~\ref{lemma:decomposition_main}]
Fix a completed request \(q\) of commodity \((c,\phi)\). Let \(L^q_{uv}\) and \(L^q_u\) denote the communication and computation workloads of \(q\) served before completion. Since every service function is executed exactly once, the realized execution defines a node map \(m^q_{\mathcal V}\), with auxiliary sources and sinks mapped to their fixed network endpoints. Every valid mapping with this node map has the same node computation load \(L^q_u\).

For each augmented dependency \(e\), trace the copy of every required packet that is consumed by the downstream function or delivered to the auxiliary sink, and remove cycles from the resulting walk. This gives a simple path between the prescribed endpoints, or the empty path for co-located endpoints. Let \(w^q_{e,p}\) be the workload assigned to path \(p\). Then
\[
    \sum_p w^q_{e,p}=\xi^e,
    \qquad
    \sum_{e\in\bar{\mathcal E}^{\phi}}\sum_p
    w^q_{e,p}\mathbb I\{(u,v)\in p\}
    \le L^q_{uv}.
\]

Select one path for each dependency with product weight
\[
    \omega^q(\mathbf p)
    =
    \prod_{e\in\bar{\mathcal E}^{\phi}}
    \frac{w^q_{e,p_e}}{\xi^e}.
\]

Together with \(m^q_{\mathcal V}\), every tuple with positive weight defines a valid mapping. Let \(\theta^q_k\) be the resulting weight of mapping \(k\). The weights sum to one and have path marginals \(w^q_{e,p}/\xi^e\); hence
\[
\begin{aligned}
    \sum_k\theta^q_k\mathtt M^{(c,\phi)}_{k,uv}
    &=
    \sum_{e,p}w^q_{e,p}\mathbb I\{(u,v)\in p\}
    \le L^q_{uv},\\
    \sum_k\theta^q_k\mathtt M^{(c,\phi)}_{k,u}
    &=L^q_u.
\end{aligned}
\]

\end{proof}

The path reduction follows the efficient route idea used in UCNC~\cite{UCNC}. The additional issue for service DAGs is that the path choices of different dependencies must remain consistent with the same function placements of a request; the product construction above preserves this consistency.

\subsection{Necessity of the LP Characterization}

\begin{proof}[Proof of necessity in Theorem~\ref{theorem:capacity}]
Fix an admissible policy \(\pi\) that supports \(\boldsymbol\Lambda\), on a sample path where the support condition and the arrival and wireless state strong laws hold. The support condition and the arrival strong law imply that, for every commodity,
\[
    \lim_{T\to\infty}
    \frac{\kappa^{(c,\phi),\pi}(T)}{T}
    = \lambda^{(c,\phi)}.
\]
Define the empirical mapping rate for commodity \((c,\phi)\) by
\[
    f_k^{(c,\phi)}(T)
    =
    \frac{1}{T}
    \sum_{\substack{q\text{ of commodity }(c,\phi)\\
                    \text{completed by }T}}
    \theta^q_k.
\]

Then \(\lim_{T\to\infty}\sum_k f_k^{(c,\phi)}(T)=\lambda^{(c,\phi)}\). Summing Lemma~\ref{lemma:decomposition_main} over completed requests and using the physical resource constraints gives
\[
\begin{aligned}
    \sum_{(c,\phi),k}
    f_k^{(c,\phi)}(T)\mathtt M^{(c,\phi)}_{k,uv}
    &\le
    \frac{1}{T}\sum_{t=0}^{T-1}
    \mu_{uv}\sigma_{uv}(t)s^\pi_{uv}(t),\\
    \sum_{(c,\phi),k}
    f_k^{(c,\phi)}(T)\mathtt M^{(c,\phi)}_{k,u}
    &\le \mu_u.
\end{aligned}
\]

The empirical mapping rates and the joint frequencies of wireless states and activations are bounded, and there are finitely many of them. Hence, we can choose a common subsequence of time horizons along which they all converge. Let \(f_k^{(c,\phi)}\) denote the limiting mapping rates. For each state \(\sigma\) with \(p_\sigma>0\), let \(\alpha_{\sigma,s}\) denote the limiting fraction of those slots in which activation \(s\) is selected. Set \(\alpha_{\sigma,s}=0\) when \(p_\sigma=0\). By the strong law, state \(\sigma\) occurs with limiting frequency \(p_\sigma\). Since exactly one feasible activation is selected in each slot,
\[
    \sum_{s\in\ActSet(\sigma)}\alpha_{\sigma,s}=1,
    \qquad p_\sigma>0.
\]

The resulting average link service vector belongs to \(\overline{\mathcal C}\). Passing to the limit along the common subsequence gives all flow and resource constraints defining \(\Gamma_{\rm LP}\). Hence \(\boldsymbol\Lambda\in\Gamma_{\rm LP}\), which proves the necessity direction.
\end{proof}

\section{Additive Mapping Gap and Virtual Queue Stability}\label{app:virtual}

Throughout this appendix, \(r\) indexes a link or computation node, and \(\mathtt M^{(c,\phi)}_{k,r}\) denotes the corresponding mapping load.

\subsection{Proof of Proposition~\ref{prop:fixedK_additive_gap}}

\begin{proof}
Let
\[
    R_{\max}^{(c,\phi)}
    =
    \max_k\sum_r \mathtt M^{(c,\phi)}_{k,r}
    <\infty.
\]

For \(\FrameLen=1\), the mapping uses the current prices and the gap is zero. For \(\FrameLen>1\), Assumption~\ref{assumption:fixedK_bounded_arrival} and the virtual queue recursions give
\[
\begin{aligned}
\left|\widetilde Q_r(t+1)-\widetilde Q_r(t)\right|
&\le\Delta_{\max},\\
\Delta_{\max}
&=\sum_{(c,\phi)}A_{\max}^{(c,\phi)}R_{\max}^{(c,\phi)}+\mu_{\max},
\end{aligned}
\]
where \(\mu_{\max}\) is the largest node or link capacity. Thus, for \(t\in\mathcal T_n\),
\[
    \left|\widetilde Q_r(t)-\Qhat_r(n)\right|
    \le
    G_{\FrameLen},
    \qquad
    G_{\FrameLen}=(\FrameLen-1)\Delta_{\max}.
\]

If \(\widehat\Psi_n^{(c,\phi)}(k)\) denotes the mapping cost under the frame head prices, then
\[
    \left|
    \Psi_t^{(c,\phi)}(k)
    -
    \widehat\Psi_n^{(c,\phi)}(k)
    \right|
    \le
    G_{\FrameLen}R_{\max}^{(c,\phi)}.
\]

Applying the preceding bound to the stored mapping gives
\[
\begin{aligned}
    \Psi_t^{(c,\phi)}
    \!\left(\widehat k_n^{(c,\phi)}\right)
    \le{}&
    \widehat\Psi_n^{(c,\phi)}
    \!\left(\widehat k_n^{(c,\phi)}\right)\\
    &+G_{\FrameLen}R_{\max}^{(c,\phi)}\\
    ={}&\min_k\widehat\Psi_n^{(c,\phi)}(k)
    +G_{\FrameLen}R_{\max}^{(c,\phi)},
\end{aligned}
\]
where the equality follows from the optimality of the stored mapping under the frame-head prices. Applying the same bound to every valid mapping and taking the minimum gives
\[
    \min_k\widehat\Psi_n^{(c,\phi)}(k)
    \le
    \min_k\Psi_t^{(c,\phi)}(k)
    +G_{\FrameLen}R_{\max}^{(c,\phi)}.
\]

Combining the two bounds yields
\[
    \Psi_t^{(c,\phi)}
    \!\left(\widehat k_n^{(c,\phi)}\right)
    \le
    \min_k\Psi_t^{(c,\phi)}(k)
    +2G_{\FrameLen}R_{\max}^{(c,\phi)}.
\]

Multiplying by \(A^{(c,\phi)}(t)\le A_{\max}^{(c,\phi)}\) proves Proposition~\ref{prop:fixedK_additive_gap} with
\[
    C_{\FrameLen}^{(c,\phi)}
    =
    2A_{\max}^{(c,\phi)}
    G_{\FrameLen}R_{\max}^{(c,\phi)}.
\]

In particular, \(C_1^{(c,\phi)}=0\) and \(C_{\FrameLen}^{(c,\phi)}=O(\FrameLen-1)\).
\end{proof}

\subsection{Proof of Theorem~\ref{theorem:virtual_stable}}

We use a quadratic Lyapunov drift argument, with Proposition~\ref{prop:fixedK_additive_gap} accounting for the additional cost of frame-based mapping.

Let \(\mathcal R_+\) be the resources used by at least one valid mapping, and write
\[
    \operatorname{load}_r(f)
    =
    \sum_{(c,\phi),k}
    f_k^{(c,\phi)}\mathtt M^{(c,\phi)}_{k,r}.
\]
Queues outside \(\mathcal R_+\) receive no virtual arrivals and are nonincreasing.

\emph{Uniform slack.}
For every \(\boldsymbol\Lambda\in\mathrm{int}(\Gamma_{\rm LP})\), there exist feasible mapping rates \(f\), an average service vector \(\bar{\mathbf c}\in\overline{\mathcal C}\), and \(\epsilon>0\) such that \(\sum_k f_k^{(c,\phi)}=\lambda^{(c,\phi)}\) and
\[
    \operatorname{load}_r(f)
    \le
    \bar s_r-\epsilon,
    \qquad r\in\mathcal R_+,
\]
where \(\bar s_{uv}=\bar c_{uv}\) and \(\bar s_u=\mu_u\). Since \(\boldsymbol\Lambda\) is an interior point, a slightly larger arrival rate vector is feasible. Scaling the corresponding mapping rates down creates positive slack on resources with nonzero load. By convexity of \(\overline{\mathcal C}\), positive service can also be allocated to the remaining usable links without losing this slack. Since \(\mathcal R_+\) is finite, a common \(\epsilon>0\) exists.

Condition on the history before the current arrivals and wireless state are revealed, and write \(\mathbb E_t[\cdot]\) for the corresponding conditional expectation. Let
\[
    L(t)=\sum_{r\in\mathcal R_+}\widetilde Q_r^2(t).
\]

Finite arrival second moments and bounded mapping loads give a finite drift constant \(B\). A randomized mapping that selects \(k\) with probability \(f_k^{(c,\phi)}/\lambda^{(c,\phi)}\) has mean resource load \(\operatorname{load}_r(f)\). Proposition~\ref{prop:fixedK_additive_gap} bounds the queue-weighted arrival cost of the frame policy by that of this comparison policy plus \(C_{\FrameLen}^{\rm slot}=\sum_{(c,\phi)}C_{\FrameLen}^{(c,\phi)}\), while max-weight activation provides at least the queue-weighted link service of \(\bar{\mathbf c}\). Therefore,
\begin{equation}
\begin{aligned}
\mathbb E_t[L(t+1)-L(t)]
&\le B+2C_{\FrameLen}^{\rm slot}\\
&\quad+2\sum_{r\in\mathcal R_+}\widetilde Q_r(t)
\bigl(\operatorname{load}_r(f)-\bar s_r\bigr)\\
&\le B+2C_{\FrameLen}^{\rm slot}
-2\epsilon\sum_{r\in\mathcal R_+}\widetilde Q_r(t).
\nonumber
\end{aligned}
\end{equation}

Taking expectations and telescoping yields
\[
    \frac{1}{T}\sum_{t=0}^{T-1}
    \mathbb E\!\left[
    \sum_{r\in\mathcal R_+}\widetilde Q_r(t)
    \right]
    \le
    \frac{B+2C_{\FrameLen}^{\rm slot}}{2\epsilon}
    +
    \frac{\mathbb E[L(0)]}{2\epsilon T}.
\]

Hence the active virtual queues are strongly stable. This proves \mbox{Theorem~\ref{theorem:virtual_stable}}.
\unskip\nobreak\hspace{0.5em}\hfill\(\square\)

\section{Physical Network Stability}\label{app:physical}

Under Assumption~\ref{assumption:fixedK_bounded_arrival}, virtual queue arrivals and service are bounded. Theorem~\ref{theorem:virtual_stable} and the standard implication from strong stability to rate stability~\cite{UMW,NeelySNO} therefore give, with probability one,
\[
    \lim_{t\to\infty}
    \frac{\widetilde Q_r(t)}{t}=0
    \qquad\text{for all virtual resources }r.
\]

Fix such a sample path and write \(\pi=\policyvk\). We use the hop grades and slot timing of Section~\ref{subsec:DCNC_DAG}.

\subsection{Notation and Virtual Interval Bound}\label{app:virtual_interval}

For a link, define the cumulative service allocated over \([t_0,t)\) by
\[
    C_{uv}^{\pi}(t_0,t)
    \triangleq
    \sum_{\tau=t_0}^{t-1}
    \mu_{uv}(\tau)s_{uv}^{\pi}(\tau),
\]
and for a node let \(C_u^{\pi}(t_0,t)=\mu_u(t-t_0)\). Let \(\widetilde A_r^{\pi}(t_0,t)\) be the cumulative virtual queue arrivals to resource \(r\). The running maximum
\[
    M_0(t)
    \triangleq
    \max_{0\le s\le t}\max_r\widetilde Q_r(s)
\]
is \(o(t)\). Summing the virtual queue recursion gives, uniformly over \(0\le t_0\le t\),
\begin{equation}
\begin{aligned}
    \widetilde A_r^{\pi}(t_0,t)
    &\le
    C_r^{\pi}(t_0,t)
    +\widetilde Q_r(t)-\widetilde Q_r(t_0)\\
    &\le
    C_r^{\pi}(t_0,t)+M_0(t).
\end{aligned}
\label{eq:app_virtual_interval}
\end{equation}

Let \(B_{\rm req}\) uniformly bound the total number of mapped packet transmissions, processing jobs, waiting components, and packets stored in waiting components of any one request. Such a bound exists because the commodity and valid mapping sets are finite. Each packet transmission has unit workload, and the finitely many positive function workloads are uniformly bounded above and away from zero. Thus, each unstarted packet transmission or processing job has a uniform positive workload lower bound. At any slot boundary, each link or computation node has at most one partially served object. Hence, for a finite constant \(\eta_{\rm N}\), the number \(N_{\le h}(t)\) of packets with unfinished transmissions and unfinished processing jobs of grade at most \(h\) satisfies
\begin{equation}
    N_{\le h}(t)
    \le
    \eta_{\rm N}\bigl(1+Q^{\le h}(t)\bigr),
    \label{eq:app_object_count}
\end{equation}
and the same bound holds with strict grade inequalities. Since every nonempty waiting component contains at least one packet, the number of such components of grade at most \(h\) is bounded by \(W^{\le h}(t)\).

\subsection{Physical Arrival Bound}\label{app:physical_arrival}

Let \(A_{r,\le h}(t_0,t)\) be the physical workload of grade at most \(h\) that becomes eligible at resource \(r\) during \([t_0,t)\). The following bound is the counterpart of the virtual to physical workload step used in UMW~\cite{UMW} and UCNC~\cite{UCNC}, with the waiting term accounting for merge synchronization.

\begin{lemma}[Physical arrival bound]\label{lemma:app_physical_arrival}
There exists a finite constant \(\eta_{\max}\), independent of \(t_0,t,h,r\), such that
\begin{equation}
\begin{aligned}
    A_{r,\le h}(t_0,t)
    \le{}&
    \widetilde A_r^{\pi}(t_0,t)\\
    &+
    \eta_{\max}
    \bigl(1+Q^{<h}(t_0)+W^{\le h}(t_0)\bigr).
\end{aligned}
\label{eq:app_physical_arrival}
\end{equation}
\end{lemma}

\begin{proof}
Partition the physical arrivals according to whether their requests were admitted during \([t_0,t)\) or before \(t_0\). For a request admitted during the interval, the virtual system charges the complete selected mapping at admission and the physical system executes that same mapping. Its physical work released at resource \(r\) is therefore bounded by the corresponding virtual workload.

Under bounded source delays, requests admitted before \(t_0\) but not yet released into the physical system at \(t_0\) contribute only a uniformly bounded amount of additional workload, which can be absorbed into \(\eta_{\max}\).

Now consider a grade \(g\le h\) item from a request released before \(t_0\). Trace one prerequisite backward toward time \(t_0\). Along a mapped route, move to the preceding transmission; before the first link, move to the predecessor function that produces the packet. If its processing job was present and unfinished at \(t_0\), stop at that job. If the job had not formed, stop at the function's waiting component when that component was nonempty at \(t_0\), and otherwise continue through an input that was missing at \(t_0\). Stop upon reaching a packet or processing job that was present and unfinished at \(t_0\).

The trace remains within the request and terminates because mapped routes are finite and the service DAG is acyclic. Except for a uniformly bounded boundary contribution, a terminal packet or processing job has grade strictly below \(g\), whereas a terminal waiting component has grade at most \(g\). Each terminal object belongs to one request, and that request contains at most \(B_{\rm req}\) mapped work items at resource \(r\), each with uniformly bounded workload. Thus the work generated by requests admitted before \(t_0\) is at most a constant times
\[
    N_{<h}(t_0)+W^{\le h}(t_0)+1.
\]
Using \eqref{eq:app_object_count} gives \eqref{eq:app_physical_arrival}.
\end{proof}

\subsection{Proof of Lemma~\ref{lemma:witness_charging_main}}\label{app:waiting_bound}

\begin{figure}[htbp]
    \centering
    \includegraphics[width=\linewidth]{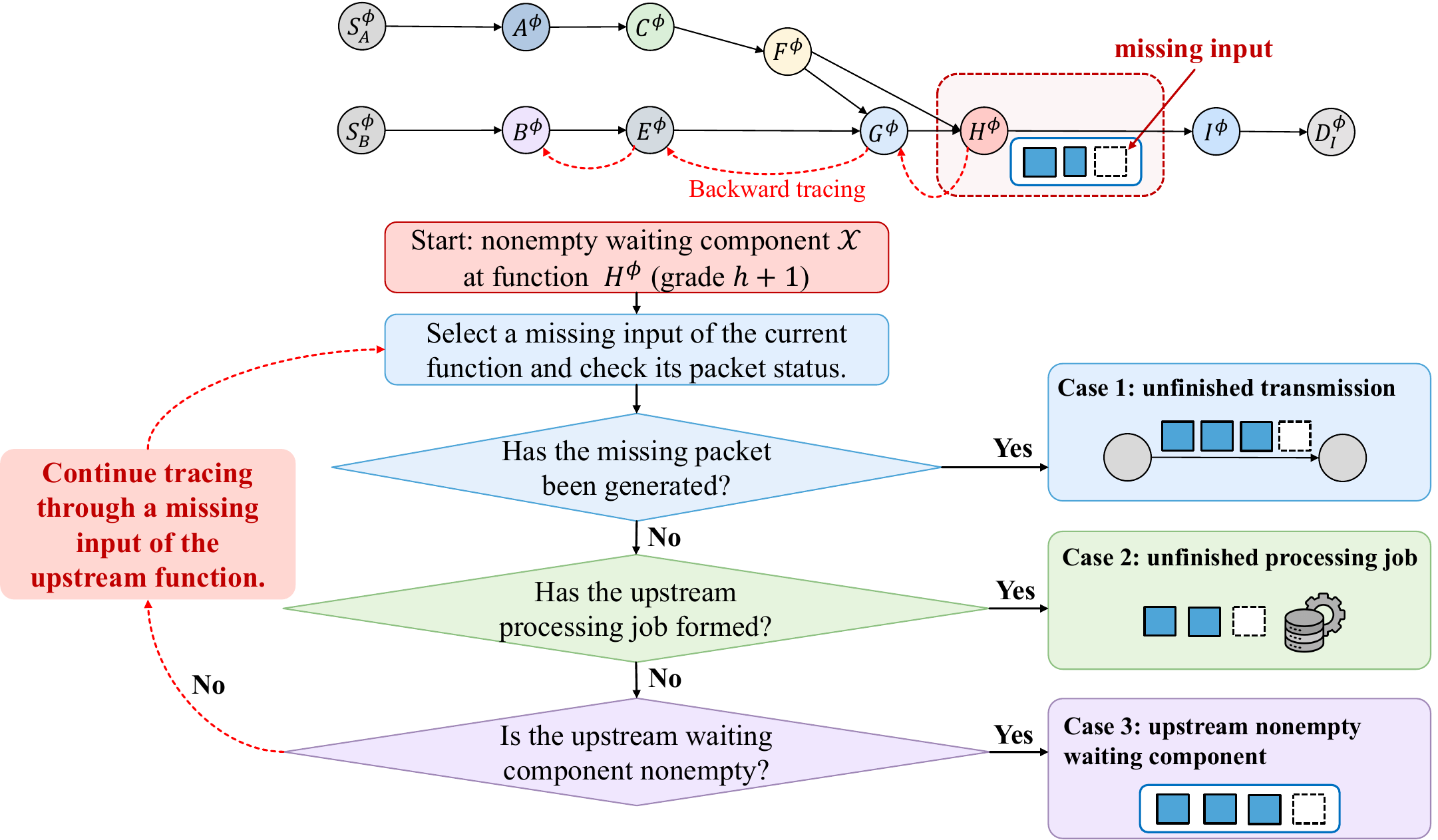}
    \caption{Backward tracing in the proof of Lemma~\ref{lemma:witness_charging_main}. Starting from a nonempty waiting component \(\mathcal X\) of grade \(h+1\), the trace follows missing inputs through predecessor functions within the same request until reaching one of the three terminal cases, each of grade at most \(h\).}
    \label{fig:app_lemma2_backward_tracing}
\end{figure}

\begin{proof}
We first associate each nonempty waiting component of grade $h+1$ with an unfinished packet, an unfinished processing job, or a nonempty upstream waiting component of grade at most $h$ (Fig.~\ref{fig:app_lemma2_backward_tracing}). We then bound the number of waiting packets associated with each such object.

Consider a nonempty waiting component $\mathcal X$ of grade $h+1$, and select one of its missing input packets. If the corresponding packet has been generated, then at least one transmission on its remaining mapped path is unfinished. The packet therefore has grade at most \(h\).

If the packet has not been generated, it must be an intermediate output, since all source packets of the request are available before its physical execution begins. Move to the predecessor function that produces this packet. If its processing job has formed, the job must remain unfinished and has grade at most \(h\), since output generation increases the grade by one. The trace then stops at that job.

If the job has not formed, its waiting component has grade at most \(h\). The trace stops if this component is nonempty; otherwise, it continues through one of the function's missing inputs. Each continuation moves to a predecessor function, so acyclicity and the availability of source packets guarantee termination in one of the three cases.

All traces remain within their originating requests. Since each terminal object belongs to one request, at most \(B_{\rm req}\) waiting packets of grade \(h+1\) can be associated with it. Traces ending at unfinished packets or processing jobs account for at most \(B_{\rm req}N_{\le h}(t)\) waiting packets. Those ending at nonempty upstream waiting components account for at most \(B_{\rm req}W^{\le h}(t)\) waiting packets, since each such component contains at least one packet. Using \eqref{eq:app_object_count}, we obtain
\[
\begin{aligned}
W^{h+1}(t)
&\le B_{\rm req}
\bigl(N_{\le h}(t)+W^{\le h}(t)\bigr)\\
&\le B_{\rm req}
\bigl[\eta_{\rm N}(1+Q^{\le h}(t))
+W^{\le h}(t)\bigr]\\
&\le \eta_{\rm w}
\bigl(1+Q^{\le h}(t)+W^{\le h}(t)\bigr),
\end{aligned}
\]
where \(\eta_{\rm w}=B_{\rm req}\max\{1,\eta_{\rm N}\}\). This proves Lemma~\ref{lemma:witness_charging_main}.
\end{proof}

\subsection{NTO Service and Hop Grade Induction}\label{app:nto_induction}

\begin{proof}[Proof of Theorem~\ref{theorem:physical_stable}]
Combining Lemma~\ref{lemma:app_physical_arrival} with \eqref{eq:app_virtual_interval} gives
\begin{equation}
\begin{aligned}
    A_{r,\le h}(t_0,t)
    \le{}&
    C_r^{\pi}(t_0,t)+M_0(t)\\
    &+
    \eta_{\max}
    \bigl(1+Q^{<h}(t_0)+W^{\le h}(t_0)\bigr).
\end{aligned}
\label{eq:app_resource_bound}
\end{equation}

Fix \(r,h,t\) with \(Q_r^{\le h}(t)>0\), and let \(\theta\) be the beginning of the slot containing the last instant before \(t\) at which this workload was empty, taking \(\theta=0\) if necessary. Thereafter, NTO starts no object of grade greater than \(h\) while work of grade at most \(h\) is waiting. At most one higher grade object can already be in service; its residual workload together with the boundary slot discrepancy is bounded by a constant \(b_r\). Queue accounting and \eqref{eq:app_resource_bound} therefore give
\begin{equation}
\begin{aligned}
    Q_r^{\le h}(t)
    \le{}&
    M_0(t)+b_r\\
    &+
    \eta_{\max}
    \bigl(1+Q^{<h}(\theta)+W^{\le h}(\theta)\bigr).
\end{aligned}
\label{eq:app_nto_recursion}
\end{equation}
The same inequality is immediate when \(Q_r^{\le h}(t)=0\).

All source packets of a request are released together into the physical system, and empty paths are resolved before service, so \(W^0(t)=0\). With \(Q^{<0}(t)=0\), \eqref{eq:app_nto_recursion} gives \(Q^0(t)=o(t)\), and Lemma~\ref{lemma:witness_charging_main} gives \(W^1(t)=o(t)\). Inductively, suppose \(Q^{<h}(t)=o(t)\) and \(W^{\le h}(t)=o(t)\). Their running maximum is sublinear, so \eqref{eq:app_nto_recursion} gives \(Q^h(t)=o(t)\); Lemma~\ref{lemma:witness_charging_main} then gives \(W^{h+1}(t)=o(t)\) whenever \(h<H_{\max}\). Since there are finitely many grades and resources, all physical queues and waiting pools are rate stable on the fixed sample path, and hence almost surely.
\end{proof}

The same physical stability argument applies with boundedly stale mapping prices, since Corollary~\ref{cor:stale} preserves virtual queue stability.

\subsection{Proof of Corollary~\ref{cor:throughput}}\label{app:throughput_proof}

\begin{proof}
Every incomplete request has an unfinished packet or processing job, or a packet in a waiting component. The preceding bounds imply that the number \(I(t)\) of incomplete requests is bounded by a constant multiple of the total unfinished physical workload and the number of packets in waiting components, plus a constant. Theorem~\ref{theorem:physical_stable} therefore implies
\[
    I(t)=o(t).
\]

For each commodity, the difference between the cumulative numbers of admitted and completed requests is at most \(I(t)=o(t)\).

Dividing by \(t\) and applying the arrival strong law gives
\[
    \lim_{t\to\infty}\frac{\kappa^{(c,\phi)}(t)}{t}
    =
    \lambda^{(c,\phi)}
    \qquad\text{w.p.\,1}.
\]

Thus every vector in \(\mathrm{int}(\Gamma_{\rm LP})\) is supportable. Together with Appendix~\ref{app:capacity}, this completes the proof of Theorem~\ref{theorem:capacity}.
\end{proof}

\section{Simulation Configuration}\label{app:simulation_config}

\subsection{Wireless Networks and Workloads}\label{app:sim_wireless}

Tables~\ref{tab:app_instances} and~\ref{tab:app_dags} specify the wireless instances and workloads under the common capacity and channel model of Section~\ref{subsec:simulation_setup}. Source and sink nodes, qualified node sets, and dynamic links are fixed for each instance and shared by all compared policies. Within each commodity, raw inputs share one source node and sink outputs share one destination node, except for the autonomous driving DAG on the Random topology, which uses two source nodes.

\begin{table}[htbp]
\centering
\caption{Wireless network simulation instances.}
\label{tab:app_instances}
\footnotesize
\setlength{\tabcolsep}{4pt}
\begin{tabularx}{\columnwidth}{@{}l L c c@{}}
\hline
Network & Service DAGs & \shortstack{Qualified\\nodes} & \shortstack{Dynamic\\links} \\
\hline
Milan & Autonomous driving & 11 & 4 \\
Random & Autonomous driving & 8 per function & 6 \\
Milan & \(D_1,D_2,D_3\) & 11 & 4 \\
Random & \(D_1,D_2,D_3\) & 16 & 6 \\
RoofNet & \(D_1,D_2,D_3\) & 10 & 35 \\
FlockLab & \(D_1,D_2,D_3\) & 10 & 66 \\
\hline
\end{tabularx}
\tabnote{Network sizes are given in Table~\ref{tab:sim_topologies}. Qualified node sets are common to all functions except for autonomous driving on the Random topology. Dynamic link counts are undirected. On RoofNet and FlockLab, all links incident to commodity endpoints are dynamic.}
\end{table}

\begin{table}[htbp]
\centering
\caption{Workloads for the service DAGs.}
\label{tab:app_dags}
\scriptsize
\setlength{\tabcolsep}{3.5pt}
\begin{tabularx}{\columnwidth}{@{}l L@{}}
\hline
Autonomous driving & \(r:\ A\,2,B\,2,C\,1,E\,3,F\,1.5,G\,1.5,H\,2\); in/out: \(A\,1.5,B\,2,C\,0.5/F\,1,H\,1\); \(\xi:\ A\!\to\!E\,1.5,B\!\to\!E\,1.5,B\!\to\!F\,1.5,C\!\to\!F\,0.5,E\!\to\!G\,2,G\!\to\!H\,1,F\!\to\!H\,1\) \\
\hline
\(D_1\) & \(r=(2,5,1,3,4,2)\); in/out: \(s_1\,4/s_6\,5\); \(\xi:\ s_1\!\to\!s_2\,3,s_1\!\to\!s_3\,5,s_2\!\to\!s_4\,2,s_3\!\to\!s_4\,4,s_3\!\to\!s_5\,1,s_4\!\to\!s_6\,5,s_5\!\to\!s_6\,2\) \\
\hline
\(D_2\) & \(r=(4,1,5,2,3,4)\); in/out: \(s_1\,2,s_2\,3/s_6\,1\); \(\xi:\ s_1\!\to\!s_3\,2,s_2\!\to\!s_3\,4,s_1\!\to\!s_4\,5,s_3\!\to\!s_5\,1,s_4\!\to\!s_5\,3,s_3\!\to\!s_6\,2,s_5\!\to\!s_6\,5\) \\
\hline
\(D_3\) & \(r=(5,2,4,1,3,5)\); in/out: \(s_1\,1/s_5\,4,s_6\,2\); \(\xi:\ s_1\!\to\!s_2\,4,s_1\!\to\!s_3\,1,s_2\!\to\!s_4\,3,s_3\!\to\!s_4\,5,s_2\!\to\!s_5\,2,s_4\!\to\!s_5\,4,s_4\!\to\!s_6\,1\) \\
\hline
\end{tabularx}
\tabnote{Letters denote service functions in the autonomous driving DAG; \(s_1,\ldots,s_6\) denote those in \(D_1,D_2,D_3\), in the order used for \(r\).}
\end{table}

\subsection{Simulation Configurations}\label{app:sim_configurations}

\begin{table}[!t]
\centering
\caption{Simulation configurations used in Section~\ref{sec:simulation}.}
\label{tab:app_configurations}
\scriptsize
\setlength{\tabcolsep}{2.2pt}
\renewcommand{\arraystretch}{1.04}
\begin{tabularx}{\columnwidth}{@{}>{\raggedright\arraybackslash}p{0.22\columnwidth}>{\raggedright\arraybackslash}p{0.27\columnwidth}>{\raggedright\arraybackslash}X@{}}
\hline
Experiment & Run configuration & Main configuration \\
\hline
LP boundary & bounded i.i.d.; \(10^5\) slots; seeds \(\{0,1,2,3,4\}\) & \(\FrameLen=1\); offered multipliers \(\rho\in\{0.95,0.98,\allowbreak 0.99,1.01,\allowbreak 1.03,1.05,\allowbreak 1.10\}\) \\
Wireless mapping & Poisson; \(2\times10^4\) slots; seeds \(\{1,2,3\}\) & policy-specific rate grids \\
Workflow DAGs & Poisson; \(5\times10^4\) slots; seeds \(\{0,1,2\}\) & workflow rate grids refined near \(\hat\lambda\) \\
MoE DAGs & Poisson; \(10^5\) slots; seeds \(\{0,1,2\}\) & MoE rate grids refined near \(\hat\lambda\) \\
Frame length & Poisson; \(10^5\) slots; seeds \(\{1,2,3\}\) & \(\FrameLen\in\{1,4,16,64,256\}\); common grid within each topology \\
FIFO / NTO & Bernoulli; \(10^7\) slots; ten seeds & \(\lambda=0.80\), empty start; same arrivals, mapping, and capacities; service order differs \\
\hline
\end{tabularx}
\end{table}


\emph{LP capacity-boundary instance.} The fork-join DAG of Section~\ref{subsec:throughput_boundary_validation} has \(r^{A^\phi}=r^{B^\phi}=r^{D^\phi}=0.1\), \(r^{C^\phi}=1\), dependency workloads \(1\) on the \(C\) branch and \(0.2\) on the \(B\) branch, and input/output workload \(0.2\). Functions \(A,B,D\) are pinned to nodes \(u_A,u_B,u_D\), while \(C\) may execute at \(u_{C1}\) or \(u_{C2}\). The two candidate nodes have capacity \(0.5\); the other three computation nodes have capacity \(5\). All links have capacity \(2\); the four links on the \(C\) branch are independently ON with probability \(0.85\), and the others remain ON under primary interference. Removing the computation constraints gives wireless capacity about \(1.43\). Arrivals are \(\lfloor\lambda\rfloor+\mathrm{Bernoulli}(\lambda-\lfloor\lambda\rfloor)\).

\begin{samepage}
\emph{Frame-length threshold and delay.} The \(10^5\)-slot runs use the following offered-rate grids for the empirical supported rates in Table~\ref{tab:sim_frame_overhead} and the delay curves in Fig.~\ref{fig:sim_frame_length}:
\[
\begin{aligned}
\text{Milan: }&\{0.049,0.1176,0.2058,0.3038,\\
&0.4116,0.5194,0.6174,0.7056,\\
&0.784,0.8428,0.8918,0.931,\\
&0.9702,1.0094,1.0584,1.1074\},\\
\text{Random: }&\{0.03,0.072,0.126,0.186,\\
&0.252,0.318,0.378,0.432,\\
&0.48,0.516,0.546,0.57,\\
&0.594,0.618,0.648,0.678\}.
\end{aligned}
\]
\end{samepage}%
Fig.~\ref{fig:sim_frame_length} shows the completed-request delay curves over the full rate grids for every frame length; the supported-rate estimates use these same grids.

\emph{Frame-length mapping overhead.} Calls and wall-clock timing use separate five-point rate grids: \(\{0.1,0.2,0.5,0.8,0.9\}\) on Milan and \(\{0.1,0.2,0.3,0.42,0.5\}\) on Random. Timing was measured on an AMD EPYC 7763 server.

\emph{FIFO/NTO instance.} The six functions visit two unit-capacity nodes \(u_1\) and \(v_1\). The first and last functions execute at \(u_1\) with workloads \(0.01\) and \(0.99\); the four intermediate functions execute at \(v_1\) with workloads \(0.99,0.01,0.01,0.01\). All links are always ON and have sufficiently large capacities so that they do not limit throughput. The per-request loads are \(1.00\) at \(u_1\) and \(1.02\) at \(v_1\), giving \(\lambda^{\star}=\min\{1/1.00,1/1.02\}=1/1.02\approx0.9804\). We use Bernoulli arrivals with \(\lambda=0.80\), start from an empty system, and simulate \(10^7\) slots over ten seeds. For each seed, FIFO and NTO use the same arrival sample path. FIFO orders jobs by their arrival time at the current node, including returning jobs, whereas NTO serves the smallest hop grade first.

Exact mappings were computed by dynamic programming, and the state-averaged LP was verified using SciPy/HiGHS.

\subsection{Large Service DAG Configuration}\label{app:sim_large}

The MoE workloads are synthetic models based on operator magnitudes, whereas the workflow DAGs are derived from published application instances.

\begin{table}[htbp]
\centering
\caption{MoE and workflow configurations used in Section~\ref{subsubsec:large_service_dags}.}
\label{tab:app_largedag}
\footnotesize
\setlength{\tabcolsep}{4pt}
\begin{tabularx}{\columnwidth}{@{}l L@{}}
\hline
\multicolumn{2}{@{}l}{MoE serving} \\
Service DAG & Mixtral: 20 functions, 27 dependencies; Llama: 25 functions, 34 dependencies \\
Operators & Mixtral 8\(\times\)7B: attention \(\approx2.2\) TFLOPs, each expert \(\approx1.45\) TFLOPs, edges \(\approx64\) MB; Llama-4-Maverick: attention \(\approx2.4\) TFLOPs, dense feedforward \(\approx1.4\) TFLOPs, each expert \(\approx0.7\) TFLOPs, edges \(64\)--\(128\) MB; FP32, 8K-token inputs \\
Weights & attention \(1.25\gamma\); expert \((0.55+3.25w_k)\gamma\); merge \(0.28\gamma\), with \(\gamma=2.00\) (Mixtral) and \(\gamma=1.84\) (Llama) \\
Network & Milan (30, 35); its 11 qualified computation nodes are treated as edge servers equipped with GPUs and are qualified for all service functions \\
Wireless & Same Milan wireless model as in Section~\ref{subsec:simulation_setup} \\
Tenants & two tenants with distinct fixed source and sink pairs; aggregate rate split \(0.60/0.40\); expert-family demand skew \(0.70/0.18/0.12\), with different families hot \\
\hline
\multicolumn{2}{@{}l}{Workflow service DAGs} \\
SRA-Search-22 & 22 functions, 30 dependencies; WfCommons \texttt{chameleon-10a} \\
SIPHT-31 & 31 functions, 34 dependencies; Pegasus SIPHT \\
Network & Random (50, 50); all 50 nodes qualified for every function \\
Workloads & published execution costs and data sizes converted to rational computation and communication workloads \\
\hline
\end{tabularx}
\end{table}

The computation requirements are proportional to the operator magnitudes in Table~\ref{tab:app_largedag} and are used without integer rounding. Here, \(k\in\{A,B,C\}\) indexes the three expert families, \(w_k\) is the fixed tenant-specific token share of family \(k\), and \(\gamma\) is the dimensionless model scale. In family order \((A,B,C)\), the code and chat tenants use \((0.70,0.18,0.12)\) and \((0.18,0.70,0.12)\) for Mixtral, and \((0.12,0.70,0.18)\) and \((0.12,0.18,0.70)\) for Llama, respectively. These shares sum to one and are used without further normalization; they also determine the workloads of the corresponding expert branches. Attention includes query, key, value, and output projections; expert costs assume uniform token splitting among activated experts. The placement rules in Section~\ref{subsubsec:large_service_dags} share the same DAGs, capacities, candidate sets, tenant arrivals, routing, and physical scheduling configuration.

\FloatBarrier
\bibliographystyle{IEEEtran}
\bibliography{reference_TMC}

\end{document}